\documentclass[
twocolumn, preprintnumbers, amsmath,amssymb,nofootinbib]{revtex4}

\usepackage{graphicx}
\usepackage{dcolumn}
\usepackage{bm}
\usepackage{amsmath,amsfonts,amssymb,slashed,upgreek,color}
\usepackage{epsfig}

\newcommand\ee{\end{equation}}
\newcommand\be{\begin{equation}}
\newcommand\eea{\end{eqnarray}}
\newcommand\bea{\begin{eqnarray}}

\newcommand\di{\partial}

\newcommand\lsim{\mathrel{\rlap{\lower4pt\hbox{\hskip1pt$\sim$}}
        \raise1pt\hbox{$<$}}}
\newcommand\gsim{\mathrel{\rlap{\lower4pt\hbox{\hskip1pt$\sim$}}
        \raise1pt\hbox{$>$}}}
        
\newcommand{\B}{\textrm{B}}
\newcommand{\F}{\textrm{F}} 
\newcommand{\Sy}{\textrm{S}}
\newcommand{\A}{\textrm{A}}
\newcommand{\bn}{\hat{\mathbf{n}}}
\newcommand{\bN}{\hat{\mathbf{N}}}
\newcommand{\bV}{\mathbf{V}} 
\newcommand{\bx}{\mathbf{x}}
\newcommand{\bq}{\mathbf{q}}
\newcommand{\bk}{\mathbf{k}}
\newcommand{\hk}{\hat{\mathbf{k}}}
\newcommand{\Dst}{\Delta^\textrm{st}}
\newcommand{\Drel}{\Delta^\textrm{rel}}    
\newcommand{\Dlens}{\Delta^\textrm{lens}}   
\newcommand{\DAP}{\Delta^\textrm{AP}}
\newcommand{\xirel}{\xi^\textrm{rel}} 
\newcommand{\xist}{\xi^\textrm{st}} 
\newcommand{\xilens}{\xi^\textrm{lens}} 
\newcommand{\xiAP}{\xi^\textrm{AP}} 
\newcommand{\HH}{{\cal H}}
\newcommand{\bB}{b_\B}
\newcommand{\bF}{b_\F}

\newcommand{\de}{\delta}
\newcommand{\dd}{\partial}

\begin{document}


\title{Asymmetric galaxy correlation functions}

\author{Camille Bonvin}
\email{cbonvin@ast.cam.ac.uk}
\affiliation{Kavli Institute for Cosmology Cambridge and Institute of Astronomy, Madingley Road, Cambridge CB3 OHA, U.K.\\
and\\
DAMTP, Centre
for Mathematical Sciences, Wilberforce Road, Cambridge CB3 0WA, U.K.
}
\author{Lam Hui}
\email{lhui@astro.columbia.edu}
\affiliation{Institute for Strings, Cosmology and Astroparticle Physics
and Department of Physics, 
Columbia University, New York, NY 10027, U.S.A.}

\author{Enrique Gazta\~naga}
\email{gazta@ice.cat}
\affiliation{Institut de Ci\`encies de l'Espai (IEEC-CSIC),
F. Ci\`encies, C5 2-par,
 Bellaterra,  Barcelona 08193, Spain. \\}

\date{\today}

\begin{abstract}
We study the two-point cross-correlation function between two populations of galaxies: 
for instance a bright population and a faint population. We show that this cross-correlation is asymmetric 
under the exchange of the line-of-sight coordinate of the galaxies, i.e. that the correlation is different if
the bright galaxy is in front of, or behind, the faint galaxy.
We give an intuitive, quasi-Newtonian derivation of all the effects
that contribute to such an asymmetry in large-scale structure:
gravitational redshift, Doppler shift, lensing, light-cone, 
evolution and Alcock-Paczynski effects -- interestingly, the gravitational redshift term is
exactly canceled by some of the others, assuming geodesic motion.
Most of these effects are captured by previous calculations of general
relativistic corrections to the observed galaxy density fluctuation; the
asymmetry arises from terms that are suppressed by the ratio $({\cal H}/k)$
-- ${\cal H}$ is the Hubble constant and $k$ is the wavenumber -- which are more readily
observable than the terms suppressed by $({\cal H}/k)^2$.
Some of the contributions to the asymmetry, however, arise
from terms that are generally considered 'Newtonian'
-- the lensing and evolution -- and thus represent a contaminant in the search for general
relativistic corrections. We propose methods to disentangle these
different contributions. A simple
method reduces the contamination to a level of $\lsim 10$\% for
redshifts $z\lsim 1$.
We also clarify the relation to recent work on 
measuring gravitational redshifts by stacking clusters.
\end{abstract}

\maketitle


\section{Introduction}
\label{sec:intro}

It is often implicitly assumed that the galaxy two-point correlation function
is symmetric under the exchange of the pair of galaxies, but this needs not be the
case when the two galaxies in question are distinguished by certain
properties, such as luminosity, color or overdensity. Expressing such a
cross-correlation as $\langle \Delta_\B ({\bf x_1}) \Delta_\F ({\bf
  x_2}) \rangle$, where the subscripts B and F represent
'bright' and 'faint' galaxies, it is evident that this quantity does
not need to be symmetric under the exchange of the positions ${\bf x_1}$ and
${\bf x_2}$. This is especially so for the interchange of the
line-of-sight coordinate: it makes a difference whether the bright galaxy is in
front of, or
behind, the faint galaxy. 

Consider a measurement of the cross-correlation function:
\begin{eqnarray}
\label{xiBFmeasure}
\xi^{\B\F} 
= \sum_{ij} W_{ij} \Delta_\B^i \Delta_\F^j \, ,
\end{eqnarray}
where we imagine the survey is pixelized into cells labeled
by $i$ or $j$, with the galaxy overdensity $\Delta_\B$ and $\Delta_\F$
defined in each cell. The kernel $W_{ij}$ determines which
pairs of galaxies are counted towards the correlation function.
We are typically interested in the correlation function at some
fixed line-of-sight (chosen to lie along the z-direction, say)
and transverse separation:
\begin{eqnarray}
\Delta x_z \equiv x_z {}^\B - x_z {}^\F \quad , \quad
\Delta {\bf x_\perp} \equiv {\bf x_\perp} {}^\B - {\bf x_\perp} {}^\F \, .
\end{eqnarray}
The correlation function $\xi^{\B\F}$ can be written as the sum of
symmetric and anti-symmetric parts:
\begin{eqnarray}
\xi^{\B\F} = 
\xi^{\B\F}_\Sy + 
\xi^{\B\F}_\A \, ,
\end{eqnarray}
where
\begin{eqnarray}
&& \xi^{\B\F}_\Sy \equiv 
{1\over 2} \Big[ \xi^{\B\F} (\Delta x_z , \Delta {\bf x_\perp}) 
+ \xi^{\B\F} (-\Delta x_z , -\Delta {\bf x_\perp}) \Big] \, , 
\nonumber \\
&& \xi^{\B\F}_\A \equiv 
{1\over 2} \Big[ \xi^{\B\F} (\Delta x_z , \Delta {\bf x_\perp}) 
- \xi^{\B\F} (-\Delta x_z , -\Delta {\bf x_\perp}) \Big]  
\label{xiasymm}
\end{eqnarray}
In general, we do not expect an asymmetry associated with
flipping the sign of the transverse separation. Thus, henceforth
we think of $\xi^{\B\F}$ as a function of $\Delta x_z$ and 
$|\Delta {\bf x_\perp}|$.

It is customary in measurements of galaxy correlation functions to use a kernel $W_{ij}$ that is symmetric under the exchange of $i$ and $j$, which necessarily captures only the symmetric part $\xi^{\B\F}_\Sy$. In this paper, we are interested in the anti-symmetric part $\xi^{\B\F}_\A$. To measure it we need to be careful about the choice of kernel and use a $W_{ij}$ that is anti-symmetric under the
exchange of $i$ and $j$. Under ensemble averaging, such an anti-symmetric kernel would cancel out the symmetric part $\xi^{\B\F}_\Sy$
and isolate the contributions from
$\langle \Delta_\B^i \Delta_\F^j \rangle - \langle \Delta_\B^j
\Delta_\F^i \rangle$.
\footnote{An example is $W_{ij} \propto$:
\begin{align}
& \, \Theta (x_z {}_i - x_z {}_j \in d_z \pm
\delta d_z) \Theta (|{\bf x}_\perp {}_i - {\bf x}_\perp {}_j| \in
d_\perp \pm \delta d_\perp) \nonumber \\
& \, -\Theta (x_z {}_j - x_z {}_i \in d_z \pm
\delta d_z) \Theta (|{\bf x}_\perp {}_i - {\bf x}_\perp {}_j| \in
d_\perp \pm \delta d_\perp) \nonumber
\end{align}
where $\Theta = 1$ if $|{\bf x}_i - {\bf x}_j|$ falls within the range
of interest, and $\Theta = 0$ otherwise. Here $d_z$ and $d_\perp$ are the line-of-sight and transverse
components of the separation.} Note that here we are interested in constructing a $W_{ij}$ which is anti-symmetric
under the flipping of the line-of-sight, but not the transverse, coordinates.

The asymmetry of interest should be distinguished from
the asymmetry that exists in the more familiar case of
redshift-space distortions.
Redshift-space distortions give rise to a correlation function
that depends on the line-of-sight separation in a way that
is different from the transverse separation (i.e. this asymmetry
is often described as an {\it anisotropy} of the redshift-space correlation
function). The asymmetry
we are interested in is a cross-correlation function that depends
on the {\it sign} of the (line-of-sight) separation.

\begin{figure}[t]
\centerline{\epsfig{figure=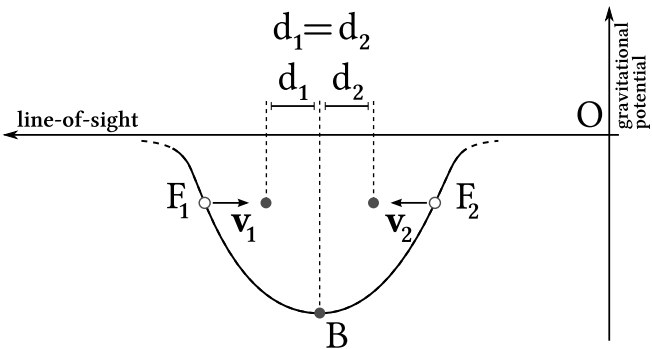,height=4.6cm}}
\caption{ \label{fig:potentiala} Sketch of the redshift-space distortion effect. The observer is sitting at O. Due to their peculiar velocities the faint galaxies $\F_1$ and $\F_2$ are shifted towards the bright galaxy and the correlation function is squeezed along the line-of-sight direction. The redshifted separation $d_1$ and $d_2$ are the same so that the effect is completely symmetric.}
\end{figure}

Why should one expect an asymmetry at all? To get an intuitive feel
for this, let us consider a highly idealized situation where we have
galaxies sitting inside the symmetric gravitational potential well
of a cluster (see Fig.~\ref{fig:potentiala}). 
We observe from afar (O in the figure). 
Let us denote by B the central cluster galaxy, located at the bottom of
the gravitational potential. Let us use F to label the other
cluster members. We use the subscript $1$ and $2$ to denote
two such members, one on each side of B i.e. $\F_1$ is behind
and $\F_2$ is in front of B, physically equidistant from B (in real
space). In redshift space, the relative positions
of the three galaxies are shifted. Fig.~\ref{fig:potentiala}
shows the Doppler effect due to infall: the galaxies are squeezed
closer together, but the effect is symmetric, in the sense that 
$\F_1$ and $\F_2$ remain equidistant from B. 
Virialized motions would give a stochastic shift in redshift space,
but on average, would still yield a symmetric effect meaning that it does not matter
whether F is in front of, or behind, B.

\begin{figure}[!t]
\centerline{\epsfig{figure=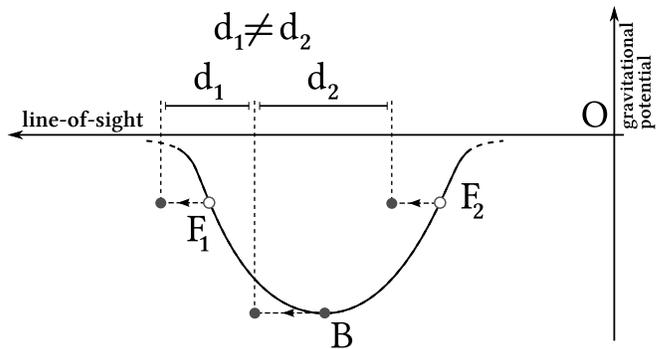,height=4.6cm}}
\caption{ \label{fig:potentialb} Sketch of the gravitational redshift
  effect. The observer is sitting at O. Galaxy B suffers the largest
gravitational redshift because it is sitting at the bottom of the
potential well. $\F_1$ and $\F_2$ shifts by a somewhat smaller amount.
The net effect is an asymmetric, $d_1\neq d_2$. 
This generates an asymmetric cross-correlation
function: B is differently correlated with F galaxies behind it
than in front of it.}
\end{figure}

The situation is different when one considers the effect of
gravitational redshift, depicted in Fig. \ref{fig:potentialb}.
Here, all 3 galaxies are shifted in the same direction, but
B is suffering the largest gravitational redshift.
The net effect is {\it asymmetric}: $\F_1$ now appears closer to B
than $\F_2$ is. 

This is of course a highly idealized example, but
the basic principle is sound: gravitational redshift yields an
asymmetric effect, which one can hope to isolate from realistic 
clusters by averaging or stacking. 
This idea was carried out in a ground breaking paper by
Wojtak, Hansen \& Hjorth
\cite{nature} (WHH): by stacking $\sim 8000$ clusters, they 
detected a net blue-shift of the average of the cluster members relative to the central brightest galaxy.

From our point of view, this is essentially a cross-correlation measurement. One can see intuitively from Fig.~\ref{fig:potentialb} that the cross-correlation between B and F is different if F is behind B or if F is in front of B. As will become clear in section~\ref{sec:results}, the correlation is actually stronger for faint galaxies behind, than in front of, the bright one. This is due to the fact that bright galaxies have a larger bias than faint galaxies (see also Fig.~\ref{fig:shiftBF} in appendix~\ref{app:derivationrel}).

This cross-correlation was first studied by McDonald~\cite{mcdonald} and Yoo et al.~\cite{uros} in Fourier space, and by Croft~\cite{croft} in configuration space. The equivalent in Fourier space of the anti-symmetric correlation function defined in eq.~\eqref{xiasymm} is an imaginary power spectrum. In~\cite{mcdonald}, McDonald showed that, in large-scale structure, the gravitational redshift effect depicted in Fig.~\ref{fig:potentialb} is not the only effect contributing to the imaginary part of the power spectrum. Other relativistic effects, directly proportional to the line-of-sight galaxy peculiar velocity, contribute in a similar way. At linear order in perturbation theory, these effects have been calculated in Refs~\cite{Yoo, galaxies, challinor}. 

Our paper expands on these previous studies in three ways. First, we give an intuitive derivation of all the effects that contribute to the anti-symmetric part of the correlation function and we show how this recovers the general relativistic expression of~\cite{Yoo, galaxies, challinor}. Second, we propose methods to disentangle the anti-symmetries directly generated by the relativistic terms from anti-symmetries induced by the redshift evolution of the density and redshift-space distortion terms (that we call hereafter 'Newtonian terms'). This second class of anti-symmetries, which is not due to new physical effects, constitute a potential contaminant in the measurement of relativistic effects. We   
provide analytical expressions, in real space, of the dipolar and the octupolar modulations of the correlation function, generated by the relativistic terms and by the contaminating Newtonian terms. Based on these expressions we then propose model-independent ways to disentangle the two classes of contributions. Finally, we use our intuitive derivation to investigate possible new effects contributing to the anti-symmetric correlation function. In particular, we study how our imperfect knowledge of the background cosmology generates anti-symmetries, through the so-called Alcock-Paczynski effect~\cite{AP}. We show that this contribution is significantly suppressed with respect to the relativistic contributions.

In practice, what we need is to work out
\begin{eqnarray}
\label{dDelta}
\Delta_{\rm obs} = \Delta + \delta\Delta\, ,
\end{eqnarray}
where $\Delta_{\rm obs}$ is the observed overdensity, $\Delta$ is
the true overdensity, and $\delta\Delta$ is the difference. 
The B and F labels have been suppressed.
Schematically, the general relativistic contributions to
$\delta\Delta$
look like
\begin{eqnarray}
\label{dDeltaGR}
\delta\Delta \sim {1\over {\cal H}} \partial_{x_z} \Psi \, ,
\end{eqnarray}
where 
${\cal H}$ is the comoving Hubble parameter i.e.
${\cal H} = \dot a / a$ with $a$ being the scale factor
and $\dot{}$ denoting derivative with respect to conformal time.
That it is the {\it gradient} of the gravitational potential $\Psi$
that matters should be apparent from Fig. \ref{fig:potentialb}:
it is the difference in gravitational potential between B and F
that is observationally relevant.
The fact that it is a {\it single} gradient
is the key to why the cross-correlation function acquires
an anti-symmetric part; squaring Eq. (\ref{dDelta}), we see
that there is a cross-term of the form
\begin{eqnarray}
\langle \Delta ({\bf x}') \delta\Delta ({\bf x}) \rangle \sim  {\cal
  H}\, \partial_{x_z} 
\nabla^{-2} \langle \Delta ({\bf x}') \Delta ({\bf x}) \rangle \, ,
\end{eqnarray}
where we have used the Poisson equation to relate $\Delta$ and $\Psi$.
Clearly, whether the derivative is with respect to $x_z$ or $x'_z$ matters.
This by itself is not so interesting if we have only one population
of galaxies, in that case the two cross-terms cancel:
$\langle \Delta ({\bf x}') \delta\Delta ({\bf x}) \rangle
+ \langle \Delta ({\bf x}) \delta\Delta ({\bf x}') \rangle = 0$.
It is only when one of the cross-terms carries a B label, and the other
F, that something non-trivial remains.

Eq. (\ref{dDeltaGR}) gives only one among many general relativistic
corrections to $\Delta$. They have been systematically worked out
by a number of authors \cite{Yoo, galaxies, challinor}. 
Two comments are in order about them.
One is that the term displayed in eq. (\ref{dDeltaGR}) is in fact
exactly canceled by some other terms, assuming geodesic galaxy motion (i.e. no
equivalence principle violation such as in \cite{HNS2009}).
This will be made explicit in our quasi-Newtonian
derivation. However, there remain terms that give a similar asymmetric
effect. In fact, they turn out to arise from Doppler terms, despite
what we said about the Doppler effect in
Fig. \ref{fig:potentiala}. Precisely how this comes about will be
made clear in the next section.

The other comment is that $\delta\Delta$ 
contains a lot of terms of order $\sim \Psi$, with no derivatives.
It is thus useful to compare three kinds of
terms: $\Delta$, $\partial\Psi / {\cal H}$, and
$\Psi$.\footnote{$\delta\Delta$ of course also
contains Newtonian terms which are order $\sim \Delta$, for instance from
the classic redshift distortion: $\delta\Delta \sim \partial_{x_z} V^z
/ {\cal
  H}$, where $V^z$ is the peculiar velocity along the line of sight.}
Making use of the Poisson equation, they are respectively
$\sim \Delta$, $({\cal H}/k) \Delta$, and $({\cal H}/k)^2 \Delta$, 
where we have freely switched to Fourier space with comoving momentum
$k$. Therefore, to the extent we are interested mostly in
sub-Hubble fluctuations i.e. ${\cal H}/k$ somewhat smaller than unity, 
the $\sim ({\cal H}/k)^2 \Delta$ general relativistic corrections can be
ignored compared to the $\sim ({\cal H}/k) \Delta$ ones.
Another good reason we focus on the $({\cal H}/k)\Delta$ terms
is that only they give rise to asymmetric effects, by
virtue of carrying an odd number of gradients.
This is what allows us to separate them from the order $\Delta$ terms
which are larger, but by themselves do not effect an asymmetry.

The other class of contributions to $\delta\Delta$ we refer to
as Newtonian includes lensing and evolution effects.
We use the term Newtonian fairly loosely here: they are in a sense
contributions we have known about all along, but somehow the
fact that they give rise to asymmetric cross-correlations has not been
emphasized. For reasons that will be apparent below, the
lensing-induced asymmetry is generally small. The evolution effect
is easy to explain: $\Delta_\F$ evolves with redshift, thus whether F
is in front of, or behind, a fixed B galaxy would give different
cross-correlation even for the same physical separation.
Such an effect becomes a contaminant in the search for general
relativistic effects, although it could be interesting in its own
right.

One comment on the ordering of fluctuations before we proceed.
In this paper, we will be using linear perturbation theory
exclusively. There are non-negligible higher order effects
in the case of clusters, as pointed out by Zhao, Peacock \& Li~\cite{Zhao} and Kaiser~\cite{kaiser}. 
We are, on the other hand, primarily interested in large-scale structure on linear scales, where these higher order effects are negligible (see discussion before eq.~\eqref{nonlinear}).
Ordering in the sense of counting powers of fluctuation variables,
should not be confused with ordering in the sense of counting
powers of $({\cal H}/k)$ as explained above.
We mostly rely on context to differentiate between the two different meanings.

The rest of the paper is organized as follow: in
section~\ref{sec:derivation} we provide an intuitive, quasi-Newtonian
derivation of the
observed number density of galaxies $\Delta_{\rm obs}$ keeping terms
up to order $\HH/k$. A fully relativistic calculation can be found
in~\cite{galaxies, challinor, Yoo}. In section~\ref{sec:correlation}
we calculate the cross-correlation between two populations of galaxies
and we expand this correlation in multipoles. In
section~\ref{sec:results} we compare the different contributions in
the anti-symmetric correlation function and we discuss ways to isolate
the relativistic contribution. 
We conclude in
section~\ref{sec:conclusion}.
In appendix \ref{app:summary} we give a brief summary of
a systematic, fully general relativistic derivation, which is nicely
consistent with the intuitive derivation in the main text where they
overlap. 
Relegated to appendices \ref{app:derivationrel} \&
\ref{app:coeffst} and \ref{app:AP} are technical details of our derivation of the
correlation function and its moments.
Finally, in appendix \ref{app:cluster} we connect our
cross-correlation viewpoint to WHH's measurement
by stacking clusters.

\section{Derivation}

\label{sec:derivation}

Our goal here is to provide an intuitive derivation of all
effects, first order in perturbations, 
that contribute to an asymmetric correlation function.
The main point we wish to get across is that to the order of
${\cal H}/k$ of interest, Newtonian or quasi-Newtonian reasoning is
sufficient for grasping the relevant effects.
More careful and systematic derivations can be found in~\cite{galaxies, challinor, Yoo}, to which we
are adding a few effects that are generally ignored, or not
emphasized.
For the convenience of the reader, we summarize
the more formal general relativistic derivation in appendix~\ref{app:summary}.

In the derivation, we use the small angle approximation, meaning
that the line-of-sight direction $\hat {\bf n}$
always points approximately in the z-direction. Deviation from the
z-direction is assumed small. 
However, the final expression will turn out to be valid on the full-sky.

We begin by relating the observed galaxy overdensity to the true
galaxy overdensity; after that, we compute the cross-correlation
function between two populations of galaxies.
We use the words 'bright'~(B) and 'faint'~(F) to distinguish between
them, but they could also be distinguished by different properties,
such as color.

For any given galaxy, we can measure its redshift and angular position
on the sky. In this paper, we assume 
the correct (homogeneous) FRW background cosmology is known, so that
these can be turned into the {\it apparent} three-dimensional
comoving position ${\bf x}_{\rm obs}$. 
In reality, our knowledge of the background cosmology is imperfect
of course. However, as we will show in section~\ref{sec:AP} and~\ref{sec:APcorr} our current knowledge is sufficiently
precise that effects associated with varying the background cosmology
within experimental bounds are small -- more precisely,
their anti-symmetric contributions to the cross-correlation function
are suppressed compared to the relativistic effects of interest by 
close to two orders of magnitude.

The true position of a galaxy ${\bf x}$ differs from the apparent
position ${\bf x}_{\rm obs}$, because the universe is in fact
not exactly FRW. 
For convenience, we use a perturbed FRW metric of the form:
\begin{eqnarray}
\label{metric}
ds^2 = a^2 \left[ -(1 + 2\Psi) d\eta^2 + (1-2\Phi) |d {\bf x}|^2
\right] \, ,
\end{eqnarray}
where $a$ is the scale factor as a function of conformal time $\eta$,
and $\Psi$ and $\Phi$ are scalar fluctuations as a function of space
and time. This is in the so-called conformal Newtonian gauge.
Observable quantities are of course
gauge independent. 

Our goal is to relate the true position of a galaxy ${\bf x}$ to
the apparent position ${\bf x}_{\rm obs}$, let
\begin{eqnarray}
{\bf x}_{\rm obs} = {\bf x} + \delta {\bf x} \, .
\end{eqnarray}
Our task is to write down the different contributions to $\delta {\bf
  x}$, and derive corrections $\delta\Delta$ to the galaxy overdensity:
\begin{eqnarray}
\Delta_{\rm obs} = \Delta + \delta\Delta \, ,
\end{eqnarray}
where $\Delta_{\rm obs}$ and $\Delta$ are the observed and
true overdensity respectively. The true overdensity
needs to be defined carefully, which we will do as we go over
the different effects.

\subsection{The spatial Jacobian: Doppler/gravitational redshift
and lensing corrections}

The two densities are related by
\begin{eqnarray}
(1 + \Delta_{\rm obs}) d^3 x_{\rm obs} = (1 + \Delta) d^3 x \, .
\end{eqnarray}
At first order in perturbations,  the correction to the density is therefore given by the Jacobian
\be
\label{jac}
\delta\Delta \sim 1 - {\,\rm det.} \left[\frac{\partial x_{\rm obs}^i}{\partial x^j}\right]\, .
\ee
It is sensitive to both perturbations along the line of sight and transverse to the line of sight.

\subsubsection{Gravitational and Doppler shifts: light propagation
 effects in the line-of-sight direction}

It is a familiar fact that peculiar motion contributes to the
observed redshift, and thus affects the apparent line-of-sight
position inferred from it. Gravitational redshift has a similar
effect. Combining the two, 
\begin{eqnarray}
\label{shifts}
\delta {\bf x} = {1 \over {\cal H}} \Big[ \hat {\bf n} \cdot ({\bf V} -
  {\bf V}_0)
- (\Psi - \Psi_0) \Big]
\hat {\bf n}  \, ,
\end{eqnarray}
where $\hat {\bf n}$ is the unit vector pointing from the observer
to the (apparent) position of the galaxy, ${\bf V}$ and $\Psi$ are respectively
the peculiar motion and gravitational potential (time-time part of the
metric perturbation) at the location of the galaxy, and
${\bf V}_0$ and $\Psi_0$ represent the same at the observer's location.
The fact that there is an overall factor of $1/{\cal H}$ comes precisely from
the conversion between the observed redshift and the line-of-sight comoving position,
assuming knowledge of the background FRW cosmology.
It is important to emphasize that the expression inside $\left[
  \,\,\right]$ in eq. (\ref{shifts}) includes only some of the large-scale
structure contributions to the net redshift of the object.
However, they are sufficient to account for all effects up to order
${\cal H}/k$ when working out corrections to $\Delta$, as we will see.

Note also that in eq.~\eqref{shifts} we keep only linear contributions in the velocity and gravitational potential and we neglect higher order terms. Zhao et al.~\cite{Zhao} and Kaiser~\cite{kaiser} showed that in cluster measurements, terms quadratic in the velocity are of the same order as the gravitational redshift and cannot be neglected. Here however we are interested in large-scale structure in the linear regime, where we can show that $V^2\ll \Psi$. Indeed, using that $V\sim k/\HH\,\Psi$, we have: 
\be
\label{nonlinear}
V^2\sim \left(\frac{k}{\HH}\right)^2\Psi^2\sim \delta\cdot \Psi\ll \Psi\, ,
\ee
where in the last inequality we have used that $\delta\ll1$ in the linear regime. Eq.~\eqref{nonlinear} therefore justifies why second-order terms can be neglected in eq.~\eqref{shifts} with respect to linear relativistic terms.

\subsubsection{ Lensing: light propagation effect in the transverse direction}

Angular deflection orthogonal to the z direction is given by
the well-known expression~\cite{dod}:
\begin{eqnarray}
\delta x_\perp^i = - \int_0^r dr' (r - r') \partial^i_\perp (\Phi +
\Psi) \, .
\end{eqnarray}
This expression is not exact, but once again, is adequate within our
approximation.
Here, $r$ is the line-of-sight comoving distance, or
approximately the z-component of ${\bf x}$. 

\subsubsection{Combining the effects}

Combining the gravitational/Doppler shifts and the lensing effect into eq.~\eqref{jac}
one finds that the correction to the density is, to first order in perturbations,
\begin{eqnarray}
\label{deltaDelta}
\delta\Delta = - {1\over \cal H} \hat n^i \hat n^j \nabla_i V_j- \int_0^r dr' {(r - r')r' \over r} \nabla_\perp^2 (\Psi +
\Phi)
\nonumber \\
- \left[ {2\over r {\cal H}} 
+  {\dot {\cal H} \over {\cal H}^2} \right]
\hat n^i (V_i - V_0 {}_i) 
+ {1\over {\cal H}} \hat n^i \left[\dot V_i + \nabla_i \Psi\right] \, .
\end{eqnarray}
The first line of Eq. (\ref{deltaDelta}) contains the well-known
Kaiser~\cite{kaiserformula} and lensing corrections. The terms on the second line are subleading,
smaller than the first line by order ${\cal H}/k$. 
In working out the Jacobian, we have to take a derivative with
respect to $z$ (i.e. $r$), and it is important to keep in mind the $r$
dependence hidden in time $\eta$ (i.e. to the lowest order, $\eta = \eta_0
- r$ where $\eta_0$ is the conformal time today): 
this is the origin of the velocity terms
on the second line.

So far, we have not accounted for the fact that objects are often
selected by flux, and therefore there are further (magnification bias)
corrections. They will be included below. The lensing correction in
the first term accounts only for the fact that galaxy density is
diluted by lensing magnification, due to a purely geometrical stretching of
the apparent area on the sky.

\subsection{Connecting observables on the light-cone to observables
at equal time: light-cone and evolution corrections}

Note that the expression $\Delta + \delta \Delta$ should be understood
as being evaluated at position ${\bf x}$, and at time $\eta = \eta_0 -
r$, with $\eta_0$ being the conformal time today 
(actually, the precise $\eta$ is slightly different from $\eta_0 - r$ due to
photon propagation in a perturbed universe, but the difference would
only contribute to terms higher order in fluctuations).
It is crucial to keep in mind that the time $\eta$ is tied to the
location ${\bf x}$ (or its z component). 
In other words,
a galaxy further away also emitted the observed photon further back.

To understand how this affects the observed galaxy overdensity let us connect these to 
fluctuations at a fixed time, say $\eta_*$. 
This seems straightforward:
\begin{eqnarray}
\label{evolution}
[\Delta + \delta\Delta]_{\eta}
\sim [\Delta + \delta\Delta]_{\eta_*}
+ {\partial \over \partial \eta} [\Delta + \delta\Delta]
\Big|_{\eta_*} (\eta - \eta_*)
\end{eqnarray}
with
$\eta - \eta_*$ small compared to Hubble time
i.e. ${\cal H} (\eta - \eta_*) \ll 1$. 
The time difference $\eta - \eta_*$ can also
be thought of as the spatial difference along the line
of sight $r_* - r$ where $r_* \equiv \eta_0 - \eta_*$, and
thus the smallness of ${\cal H}(\eta - \eta_*)$ has the same
meaning as the smallness of ${\cal H}/k$. In the spirit of
keeping only terms up to $O({\cal H}/k)$,
we need only keep $O(1)$ terms in 
$(\partial/\partial\eta)[\Delta + \delta\Delta]$, meaning
only $\Delta$ and the first line contribution to 
$\delta\Delta$ (Eq.~\eqref{deltaDelta}) needs to be included
in this derivative.
We refer to the second term on the right of Eq.~(\ref{evolution})
as the {\it evolution correction}, it obviously is related to the evolution
of the fluctuations at a given spatial position ${\bf x}$.

\begin{figure}[t]
\centerline{\epsfig{figure=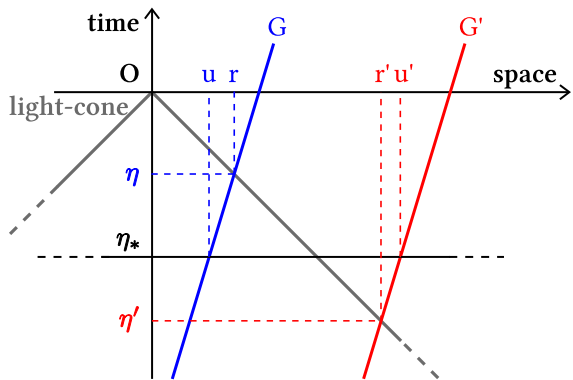,height=5cm}}
\caption{ \label{fig:lightcone} Schematic representation of the {\it light-cone effect}. Due to the peculiar velocity 
of the two galaxies $G$ and $G'$, the true line-of-sight separation between them, $r'-r$, differs
from their line-of-sight separation at the time of interest $\eta_*$: $u'-u$.}
\end{figure}

There is, however, a more subtle effect arising from the fact
that galaxies have moved between the times $\eta$ and $\eta_*$,
an effect that is not captured by differentiating the
fluctuations at a fixed ${\bf x}$. Following Kaiser~\cite{kaiser}, we
call this the {\it light-cone effect}. It is easier to explain by first
imagining that all galaxies move with the same velocity
${\bf V}$. Let ${\bf q}$ be the position of these galaxies
at the time of interest $\eta_*$. Their positions at time $\eta$
is thus: ${\bf x} = {\bf q} + {\bf V} (\eta - \eta_*)$.
On the other hand, we know $({\bf x}, \eta)$ lies on the
light-cone, meaning $r = \eta_0 - \eta$ (to lowest order);
see Fig.~\ref{fig:lightcone} for an illustration.
The line-of-sight component of ${\bf x} = {\bf q} + {\bf V}
(\eta - \eta_*)$ thus tells us
$r = u + \hat{n}^i V_i (\eta - \eta_*) = u +\hat{n}^i V_i (r_* - r)$, where $u$
denotes the line-of-sight component of $\bq$.
Solving for $r$ gives $r - r_*$ (which is
also $\eta_* - \eta$) $=(u - r_*)/(1 + \hat{n}^i V_i)
\sim u - r_*$, thus telling us $r = u + \hat{n}^i V_i (r_* - u)$,
or more generally:
\begin{eqnarray}
{\bf x} = {\bf q} + {\bf V} (r_* - u) \, .
\end{eqnarray}
This provides the mapping between the galaxy position ${\bf x}$
on the light cone and the galaxy position ${\bf q}$ at the time
of interest $\eta_*$. The Jacobian between them
gives us an additional correction
to the density, connecting the density in ${\bf x}$ space to the
density (of interest) in ${\bf q}$ space i.e.
\begin{eqnarray}
\label{lightcone}
\Delta \rightarrow \Delta + 
\left( 1 - {\rm det.}\left[{\partial x^i  \over \partial
      q^j}\right]\right)
\sim \Delta + \hat n^i V_i \, .
\end{eqnarray}
This effect can be grasped in an intuitive way from
inspecting Fig.~\ref{fig:lightcone}: due to the peculiar velocity 
of the two galaxies $G$ and $G'$, the true line-of-sight separation between them, $r'-r$, differs
from their line-of-sight separation at the time of interest $\eta_*$: $u'-u$. 
This argument assumes ${\bf V}$ has no spatial dependence.
If it does, it would seem the Jacobian should introduce
an additional correction proportional to the divergence of ${\bf V}$.
One recognizes that such a term is what contributes to
$\partial\Delta/\partial\eta$, and has already been accounted for by
the evolution corrections in Eq. (\ref{evolution}).
More precisely, if the galaxy number is conserved, we know to
linear order, $- \nabla_i V_i  = \partial\Delta/\partial\eta$, which
is contained in the corrections in Eq. (\ref{evolution}) already.
The term $\partial\Delta/\partial\eta$ could of course have additional
contributions coming from the fact that galaxies merge or form.
In any case, there is no need to account for the spatial dependence
of ${\bf V}$ when working out the light-cone effect.
Note that in the general relativistic derivation \cite{galaxies, challinor, Yoo},
the light-cone effect is automatically taken into account in the 4-dimensional space-time Jacobian.

Let us summarize, by combining Eqs.~(\ref{deltaDelta}), (\ref{evolution}) and~(\ref{lightcone}):
\begin{align}
\label{summarize0}
& \Delta_{\rm obs} = \Delta - {1\over \cal H} \hat n^i \hat n^j \nabla_i V_j
- \int_0^r dr' {(r - r')r' \over r} \nabla_\perp^2 (\Psi +
\Phi)
\nonumber \\
& \quad - \left[ {2\over r {\cal H}} 
+  {\dot {\cal H} \over {\cal H}^2} \right]
\hat n^i (V_i - V_0 {}_i) 
+ {1\over {\cal H}} \hat n^i \left[\dot V_i + \nabla_i \Psi\right]
+ \hat n^i V_i
\nonumber \\
& \quad + {\partial \over \partial \eta}
\left[\Delta 
- {1\over \cal H} \hat n^i \hat n^j \nabla_i V_j \right]
\Big|_{\eta_*}
(r_* - r) \, ,
\end{align}
where we have only kept terms up to order ${\cal H}/k$. 
Note that the time evolution correction from the lensing
term is implicit: one can expand the line-of-sight integral
$\int_0^r dr' ...$ around $r=r_*$, and the first order correction
will be proportional to $r - r_*$ just like other evolution
corrections (on the last line). 
The right hand side of Eq.~(\ref{summarize0}) should be understood
to be evaluated at position ${\bf x}$ (we need not distinguish
between ${\bf q}$ and ${\bf x}$ at this point because the difference
would contribute terms of higher order in the context of this
equation), and at time $\eta = \eta_*$. 

Eq.~\eqref{summarize0} is perfectly compatible with the general relativistic expression of~\cite{galaxies, challinor},
up to order ${\cal H}/k$ (see also eq.~\eqref{deltaobsapp} of
appendix~\ref{app:summary}). The only difference is that here the
evolution term (third line) has been made explicit, whereas
in~\cite{galaxies, challinor} it is part of the standard term
(first two terms).

\subsection{Observational selection effects}

Lastly, we wish to include three selection effects related
to observations: magnification bias, radial density variations and the Alcock-Paczynski effect.

\subsubsection{Magnification bias}

Galaxy samples are often flux limited.
This means the magnification of the flux
by intervening structure causes the apparent density
to fluctuate. The lensing term in the first line of 
Eq. (\ref{summarize0}) already accounts for the simple
geometrical stretching of the apparent area.
Here, we want to account for the fact that behind a
magnified region, faint sources that otherwise would not
have made into one's sample now do.
Previous calculations tell us that this effect gives a correction
of the form $5 s \delta_f /2$, where $\delta_f$ is the
fractional fluctuation in the flux, and $s$ is the 
effective number count slope (see e.g.~\cite{magbias, lensingHGV}):
\begin{eqnarray}
s \equiv {1\over 2.5} \left(\int df \epsilon(f) N_0 (f)  \right)^{-1}
\int df {d\epsilon \over df} f N_0 (f) \, ,
\end{eqnarray}
where $N_0 (f) df$ is the number density of sources
with flux $f \pm df/2$, and $\epsilon(f)$ describes the
detection efficiency, e.g. a step function would be an example
with a sharp flux threshold; in that case it can be shown that
$s = d{\,\rm log}_{\rm 10} N_{\rm tot}/dm$ with $N_{\rm tot}$ being the
number of galaxies brighter than magnitude $m$. 
The flux fluctuation $\delta_f$ is related to the convergence part, $\kappa$, of the magnification matrix
$\delta_f=2\kappa$ and has been calculated in~\cite{convergence, shear} 
(see also computations of the luminosity distance fluctuation: $\delta_f = - 2 \delta_{d_L}$, where $\delta_{d_L}$ is the fractional
fluctuation in luminosity distance~\cite{luminosity, HG}). The 
magnification bias correction (to be added to the right hand
side of Eq. (\ref{summarize0})) is therefore (see also~\cite{challinor}):
\begin{eqnarray}
\label{mb}
\delta\Delta_{\rm m.b.} = 
5 s \Big[ - \hat n^i V_i + {1\over r{\cal H}} \hat n^i (V_i -
V_0 {}_i) \nonumber \\ + {1\over 2} \int_0^r dr'
{(r - r') r' \over r} \nabla_\perp^2 (\Psi + \Phi) \Big]\, .
\end{eqnarray}

\subsubsection{Radial selection}

Another effect we wish to include is the fact that
in practice we often do not know the selection function
of a given galaxy survey precisely.
The correct fluctuation $\Delta$ should be 
$(n - \bar n)/\bar n$, where $n$ is the number density
and $\bar n$ its mean. In practice, we do not know
$\bar n$ precisely; suppose the assumed mean density
$\tilde n$ differs from $\bar n$ in the form
$\tilde n = \bar n (1 - \delta\Delta_{\rm sel.})$. 
It is easy to see that $\Delta_{\rm obs}$, which is
inferred from $(n - \tilde n)/\tilde n$, would contain
a correction which is $\delta\Delta_{\rm sel.}$, to lowest order.
Since the selection function is typically smooth, we can
Taylor expand $\delta\Delta_{\rm sel.}$ around the mean
redshift of the survey:
\begin{eqnarray}
\label{selection}
\delta\Delta_{\rm sel.} \sim \alpha_r (r - r_*)\, ,
\end{eqnarray}
where $\alpha_r$ is simply a constant,
describing the slope of the extent to which the assumed selection
function differs from the true selection function. We neglect here higher order terms
in the Taylor expansion, $\propto (r-r_*)^2$, which are symmetric around $r_*$ and would therefore not
contribute to the anti-symmetric correlation function~\footnote{Note that the cross-correlation between a linear term and a quadratic term would give an anti-symmetric contribution proportional to $(r-r_*)^3$. Such a contribution is however negligible as long as the redshift width of the survey is smaller than its depth $r-r_*\ll r_*$.}.  
Note that $\delta\Delta_{\rm sel.}$ could in principle contain
a constant piece as well, but such a piece is generally removed
by ensuring that the density averaged within the survey is zero.
Note also that we assume the only selection issue is in the z direction.
This is not true in general: there could well be a similar selection issue in the
angular directions as well; we will ignore this possibility.
As we will see, a term like Eq. (\ref{selection}) would not give
an interesting asymmetry in the cross-correlation function, even
though it looks like it could.

\vspace{1cm}

\subsubsection{Alcock-Paczynski effect} \label{sec:AP}

Finally, the last effect we want to discuss here is how using a (slightly) wrong background cosmology can impact the observed overdensity of galaxies $\Delta$.
We need to assume a cosmology when transforming redshift coordinates (at the level of the background) to comoving coordinates. Let us denote by $\Omega$ the true cosmology, and by $\tilde\Omega=\Omega+\delta\Omega$ the wrong cosmology, that we are using to infer $r$ from $z$. The wrong $r$ is a function of the redshift and of the wrong cosmology, $r=r(z,\tilde\Omega)$, and it is related to the true $r$ by
\be
r(z,\tilde\Omega)=r(z,\Omega+\delta\Omega)\simeq r(z,\Omega)+\frac{\partial r(z,\Omega)}{\partial \Omega}\cdot\delta\Omega\, ,
\ee
where we have neglected higher order terms in the Taylor expansion based on the fact that our knowledge of the background cosmology is relatively accurate, i.e. that $\delta\Omega/\Omega\ll 1$. Note that the direction $\bn$ of the incoming photon is not affected by the transformation from redshift to radial distances~\footnote{This however does not mean that the transverse separation between a pair of galaxies is not affected by the Alcock-Paczynski effect. On the contrary, from fig.~\ref{fig:cone} (and eq.~\eqref{rp}) we see that if $r$ and $r'$ are wrongly inferred, then the comoving separation $d$ between a pair of galaxies is also wrongly calculated.}.

The overdensity of galaxy then reads, in term of the wrong comoving coordinates:
\begin{align}
\label{AP}
\Delta(r(z,\tilde\Omega),\bn)\simeq &\,\Delta(r(z,\Omega),\bn)\\
&+\frac{d \Delta(r(z,\Omega),\bn)}{dr}
\frac{\partial r(z,\Omega)}{\partial \Omega}\cdot\delta\Omega\, .\nonumber
\end{align}
Here $d/dr$ is a total derivative, that takes into account that if $r$ is wrongly inferred from $z$, then the conformal time coordinate $\eta=\eta_0-r$ will also be wrong:
\begin{align}
\frac{d\Delta}{dr}&=\frac{\partial\Delta}{\partial r} + \frac{\dd\Delta}{\dd \eta}\frac{d\eta}{dr}
=\frac{\dd\Delta}{\dd r} - \frac{\dd\Delta}{\dd \eta}\,.
\label{Dr}
\end{align}
Combining eq.~\eqref{AP} and~\eqref{Dr} gives a correction in the overdensity, $\Delta^{\rm AP}$, of the form
\be
\label{DensityAP}
\Delta^{\rm AP}(r,\bn)=\big(\partial_r-\partial_\eta\big)\Delta(r,\bn)\frac{\partial r(z,\Omega)}{\partial \Omega}\cdot\delta\Omega\, .
\ee
The first term, proportional to $\partial_r$, brings in an additional factor $k$ and seems therefore to give an important contribution at small scales. In section~\ref{sec:APcorr} we will see however that this contribution is suppressed and that the anti-symmetric Alcock-Paczynski effect is actually smaller than the other anti-symmetric contributions by a factor $\delta\Omega/\Omega\lsim 0.01$.

\subsection{The total observed overdensity}
\label{sec:finalderivation}

Putting Eqs. (\ref{summarize0}), (\ref{mb}), (\ref{selection}) and \eqref{DensityAP}
together,
we finally obtain:
\begin{align}
& \Delta_{\rm obs} = b\cdot\delta - {1\over \cal H} \hat n^i \hat n^j \nabla_i V_j
- \int_0^r dr' {(r - r')r' \over r} \nabla_\perp^2 (\Psi +
\Phi) \nonumber \\
& \quad - \left[ {2\over r {\cal H}} 
+  {\dot {\cal H} \over {\cal H}^2} \right]
\hat n^i (V_i - V_0 {}_i) 
+ {1\over {\cal H}} \hat n^i \left[\dot V_i + \nabla_i \Psi\right]
+ \hat n^i V_i
\nonumber \\
& \quad - {\partial \over \partial \eta}
\left[b\cdot\delta 
- {1\over \cal H} \hat n^i \hat n^j \nabla_i V_j \right]
\Big|_{\eta_*}
(r - r_*) \nonumber \\
& \quad 
+5 s \Bigg[ - \hat n^i V_i + {1\over r{\cal H}} \hat n^i (V_i -
V_0 {}_i) \nonumber \\
& \quad + {1\over 2} \int_0^r dr'
{(r - r') r' \over r} \nabla_\perp^2 (\Psi + \Phi) \Bigg] \nonumber \\
& \quad + \alpha_r (r - r_*)+\DAP\, , \label{Deltagen}
\end{align}
where we have related the true galaxy overdensity $\Delta$ to the matter density contrast\footnote{Here $\delta$ denotes the density contrast in the comoving gauge, where the simple linear relation between the galaxy and matter density is valid~\cite{biasbaldauf, biasjeong}. Note however that the differences between the different gauges are of order $\HH^2/k^2$, i.e. negligible for us.} $\delta$ with the bias $b$. 
Eq.~\eqref{Deltagen} is perfectly compatible with the general relativistic result of~\cite{galaxies, challinor},
up to order ${\cal H}/k$. Our derivation provides therefore an intuitive explanation of the ${\cal H}/k$-terms contributing to the relativistic correlation function. To that we are adding two selection effects: the radial selection effect and the Alcock-Paczynski effect, in the last line of eq.~\eqref{Deltagen}.

The goal of our paper is now to identify the terms that would give rise
to an asymmetry in the cross-correlation function i.e.
squaring the above but keeping track of the fact that
there are two populations of galaxies.
As explained in section \ref{sec:intro}, terms that
turn out to contain a single gradient of the potential
give rise to an asymmetry: these are terms on the second
and fourth line of Eq. (\ref{Deltagen}), as well as some of the terms in $\DAP$.
Evolution terms, which scale as $r - r_*$, also give
rise to an asymmetry -- they are those on the third line;
there are also contributions from expanding the lensing terms on
the first and fifth lines around $r \sim r_*$.
The first term in the last line -- the radial selection effect -- does
not give anything interesting, since its only possible
anti-symmetric contribution to the cross-correlation
function comes from something
like $\langle b \delta \alpha_r (r-r_*) \rangle$ which
vanishes because $\alpha_r$ is not a fluctuating variable.
We will henceforth drop this selection effect term from $\Delta_{\rm obs}$.

\section{Correlations between two populations}

\label{sec:correlation}

Let us now study in detail the cross-correlation of $\Delta_{\rm obs}$
between two different populations of galaxies, that we label by $\B$
for bright galaxies and $\F$ for faint galaxies, but F and B  could
also refer to other selection criterion, such as color.
 In the following we ignore the contributions of the observer velocity
 $\bV_0$ that only contributes to a global dipole around the observer,
 and we ignore the selection effect in the last line of 
eq.~\eqref{Deltagen} that averages to zero when correlated with any
stochastic variable. 
For convenience we also do not write explicitly the evolution effect
 (third line of eq.~\eqref{Deltagen}): we simply absorb it into the
 density  and redshift-space distortion terms (first two terms on the
first line); the two terms should then be thought of as
being evaluated at time $\eta = \eta_0 - r$, instead of at some
fixed fiducial time $\eta_*$.
We will make the evolution terms  explicit 
again in section~\ref{sec:standard}. With this, eq.~\eqref{Deltagen} becomes
\begin{align}
& \Delta_{\rm obs} = b\cdot\delta - {1\over \cal H} \di_r (\bV\cdot\bn)
- \int_0^r d\tilde{r} {(r - \tilde{r})\tilde{r} \over r} \nabla_\perp^2 (\Psi +
\Phi) \nonumber \\
& - \left[ {2\over r {\cal H}} 
+  {\dot {\cal H} \over {\cal H}^2}-1 \right]
\bV\cdot \bn + {1\over {\cal H}} \left[\dot \bV\cdot\bn + \partial_r \Psi\right]\nonumber\\
&
+5 s \Bigg[ \Bigg({1\over r{\cal H}} -1 \Bigg)\bV\cdot\bn + \int_0^r d\tilde{r}
{(r - \tilde{r}) \tilde{r} \over 2r} \nabla_\perp^2 (\Psi + \Phi) \Bigg]\nonumber\\
&+\DAP \, .\label{Deltafin}
\end{align}
It is convenient to group the terms in eq.~\eqref{Deltafin} into four contributions: the standard contribution, the relativistic contribution, the lensing contribution and the Alcock-Paczynski contribution. With this the overdensity of bright galaxies simply reads
\bea
\Delta_\B(z,\bn)&=&\Dst_\B(z,\bn)+\Drel_\B(z,\bn)\nonumber\\
&+&\Dlens_\B(z,\bn)+ \DAP_\B(z,\bn)\, ,
\eea
where
\begin{align}
&\Dst_\B(z,\bn)=b_\B \delta(z,\bn)-\frac{1}{\HH}\di_r (\bV\cdot\bn)\label{Dst}\, ,\\
&\Drel_\B(z,\bn)=\frac{1}{\HH}\di_r\Psi + \frac{1}{\HH}\dot{\bV}\cdot\bn\label{Drelgen}\\
&\hspace{1.5cm}-\left[\frac{\dot{\HH}}{\HH^2}+\frac{2}{r\HH} -1+ 5s_\B\left(1-\frac{1}{r\HH}\right)\right]\bV\cdot\bn\, ,\nonumber\\
&\Dlens_\B(z,\bn)= (5s_\B-2)\int_0^r d\tilde{r}\frac{(r-\tilde{r})\tilde{r}}{2 r}\nabla_\perp^2 (\Phi+\Psi)\, , \label{Dlens}\\
& \DAP_\B(z,\bn)=\big(\partial_r-\partial_\eta\big)\Big(\Dst_\B+\Drel_\B+\Dlens_\B\Big)\nonumber\\
&\hspace{2.2cm}\cdot\, \frac{\partial r(z,\Omega)}{\partial \Omega}\cdot\delta\Omega\, . \label{DAP}
\end{align}
The quantity $\bB$ denotes the bias of the bright galaxies and $s_\B$ is their effective number count slope. 
{\it Expressions~\eqref{Dst} to~\eqref{DAP} are valid in any theory in
which photons travel on null geodesics; no assumptions have been
made about the dynamics of gravity.} In theories in which the
galaxies (i.e. non-relativistic tracers) also move on geodesics,
we can use the Euler equation
\be
\dot{\bV}\cdot\bn+\HH \bV\cdot\bn+\di_r\Psi=0\, ,
\ee
to simplify eq.~\eqref{Drelgen} to
\be
\Drel_\B(z,\bn)=\!-\!\left[\frac{\dot{\HH}}{\HH^2}+\frac{2}{r\HH} +5s_\B\left(1-\frac{1}{r\HH}\right) \right]\bV\cdot\bn\, . \label{Drel}
\ee
The gravitational redshift effect, $\partial_r\Psi/\HH$, is therefore canceled by a combination of the light-cone effect and part of the Doppler effect.

\begin{figure}[t]
\centerline{\epsfig{figure=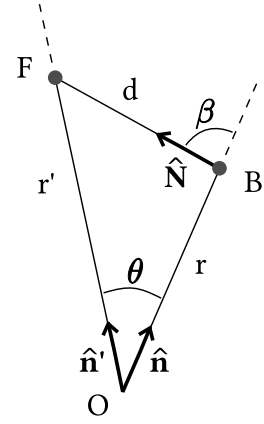,height=5.5cm}}
\caption{ \label{fig:cone} Schematic representation of the position of the bright galaxy (B) and faint galaxy (F) with respect to the observer~(O).}
\end{figure}

The overdensity of faint galaxies can be split in a similar way into a standard, relativistic, lensing and Alcock-Paczynski contribution. These contributions differ from the bright population through the bias $\bF$ and the effective number count slope $s_\F$. We can then compute the cross-correlation function $\xi^{\B\F}$ between the two populations
\be
\xi^{\B\F}(z, z', \theta)=\langle \Delta_\B(z,\bn)\Delta_\F(z',\bn') \rangle\, .
\ee
Due to statistical homogeneity and isotropy $\xi^{\B\F}$ depends only on the redshift of the bright galaxy $z$, the redshift of the faint galaxy $z'$ and the angular separation between the bright and the faint, that we denote by $\theta$~\footnote{It is worth underlying that here $z$, $z'$ and $\theta$ are the \textit{observed} redshifts and angular separation. However the important point is that we are now allowed to relate these observed quantities to the comoving radial coordinates $r$ and $r'$ and to the transverse separation $x_\perp$, using \textit{background} expressions with the \textit{wrong} cosmology. All the effects arising from the fact that fluctuations in the observed redshift and the use of a wrong cosmology modify the comoving distances have already been consistently included in the observed $\Delta$ up to linear order.} (see Fig.~\ref{fig:cone}). $\xi^{\B\F}$ contains four contributions
\begin{align}
\xist(z, z', \theta)&=\langle \Dst_\B(z,\bn)\Dst_\F(z',\bn') \rangle\, ,\\
\xirel(z, z', \theta)&=\langle \Dst_\B(z,\bn)\Drel_\F(z',\bn') \rangle\label{xirel}\\
&+\langle \Drel_\B(z,\bn)\Dst_\F(z',\bn') \rangle\, ,\nonumber\\
\xilens(z, z', \theta)&=\langle \Dst_\B(z,\bn)\Dlens_\F(z',\bn') \rangle\label{xilens}\\
+\langle \Dlens_\B(z,&\bn)\Dst_\F(z',\bn')\rangle+ \langle \Dlens_\B(z,\bn)\Dlens_\F(z',\bn')\rangle\, , \nonumber\\
\xiAP(z, z', \theta)&=\langle \big(\Dst_\B+\Drel_\B+\Dlens_\B\big)(z, \bn) \DAP_\F(z',\bn')\rangle\nonumber\\
&\hspace{-0.8cm}+\langle\DAP_\B(z, \bn)  (\Dst_\F+\Drel_\F+\Dlens_\F) (z',\bn')\rangle\, . \label{xiAP}
\end{align}
Note that in Eq.~\eqref{xirel} we have neglected the correlation between the relativistic terms $\langle \Drel_\B(z,\bn)\Drel_\F(z',\bn') \rangle$ since it is a factor $\HH/k$ smaller than the other two terms. We also neglect the correlation between the relativistic term and the lensing term that, as will become clear in section~\ref{sec:lensing}, are subdominant both in eq.~\eqref{xirel} and in eq.~\eqref{xilens}.
Finally in eq.~\eqref{xiAP} we neglect the contribution $\langle \DAP_\B(z,\bn)\DAP_\F(z',\bn') \rangle$ since it is quadratic in $\delta\Omega/\Omega\ll 1$.

We now want to calculate the anti-symmetric part of $\xi^{\B\F}$ that, as we will show hereafter, provides a way of isolating the relativistic part of the correlation function, $\xirel$. Since we have split galaxies into two populations, we are able to differentiate faint galaxies that are in front of a bright galaxy, from faint galaxies that are behind a bright galaxy. One simple way to exploit this information is to expand the two-point correlation function in multipoles around the bright galaxies. More precisely, denoting by $\beta$ the angular position of the faint galaxy with respect to the bright one (see Fig.~\ref{fig:cone}) we can expand the correlation function in multipoles of $\beta$. Even multipoles will contribute to the symmetric part of $\xi^{\B\F}$ whereas odd multipoles will contribute to the anti-symmetric part.

\subsection{Relativistic contribution}
\label{sec:relativistic}

Let us start by calculating the relativistic contribution $\xirel$ in the full-sky regime. Following refs.~\cite{szalay, szapudi, szapudi2, francesco, bertacca} we expand $\xirel$ in tri-polar spherical harmonics. We introduce a coordinate system where the triangle formed by the observer, the bright galaxy and the faint galaxy lies in the $(x_1-x_2)$ plane, see Fig.~\ref{fig:coordinate}. The unit vector $\bN$ connecting the bright to the faint galaxy is chosen to be aligned with the $x_2$ axis. In this coordinate system the directions $\bn$, $\bn'$ and $\bN$ all have a polar angle $\tilde{\theta}=\frac{\pi}{2}$. The azimuthal angle of $\bN$ is zero and we denote by $\alpha$ the azimuthal angle of $\bn'$ and $\beta$ the azimuthal angle of $\bn$. Note that $\beta$ is also the angle between the directions $\bn$ and $\bN$: it is therefore the angle in which we want to perform our multipole expansion. The coordinate $r=\eta_0-\eta$, denotes the radial comoving distance of the bright galaxy, $r'=\eta_0-\eta'$ the radial comoving distance of the faint galaxy, and $d$ is the comoving distance between the bright and the faint galaxies. Our goal is to express the correlation function in terms of this new coordinate system, more precisely as a function of $r$ (or $\eta$), $d$ and $\beta$. Note that here we consider $\eta$ as an 'observable' coordinate, in the sense that we can relate it to the measured redshift $z$ using the background relation between redshift and comoving time. The perturbations of the redshift have been consistently included in the derivation of the observed galaxy overdensity.  

A detailed derivation of the two-point function $\xirel$ is given in appendix~\ref{app:derivationrel}. Here we simply write the final result:

\begin{figure}[t]
\centerline{\epsfig{figure=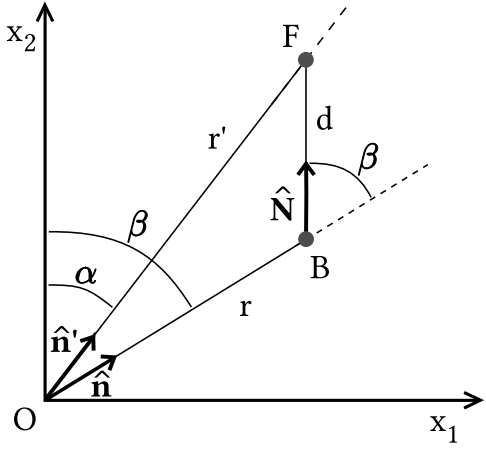,height=5.9cm}}
\caption{ \label{fig:coordinate} Definition of our coordinate system. The triangle composed of the observer (O), the bright galaxy (B) and the faint galaxy (F) lies in the $(x_1-x_2)$ plane. The unit vector $\bN$ relating the bright to the faint galaxy is parallel to the axis $x_2$.
In this coordinate system, the azimuthal angle of the bright galaxy $\beta$ is also the angle between $\bn$ and $\bN$.}
\end{figure}

\bea
\xirel&=&\frac{2A}{9\Omega_m^2\pi^2}\Bigg\{R_1 \cos(\alpha)+R_2\cos(\beta)\label{xirelalpha}\\
&&+R_3\cos(\alpha)\cos(2\beta)+R_4\cos(\beta)\cos(2\alpha)\nonumber\\
&&+R_5\sin(\alpha)\sin(2\beta)+R_6\sin(\beta)\sin(2\alpha)\Bigg\}\, ,\nonumber
\eea 
where the coefficients $R_1$ to $R_6$ are given in appendix~\ref{app:derivationrel}. These coefficients depend on $r, \;r'$ and $d$. 
We can then explicitly rewrite eq.~\eqref{xirelalpha} in terms of $r, d$ and $\beta$ using the trigonometric relations
\bea
r'&=&\sqrt{r^2+d^2+2dr\cos(\beta)}\label{rp}\, ,\\
\cos(\alpha)&=&\frac{d+r\cos(\beta)}{\sqrt{r^2+d^2+2dr\cos(\beta)}}\label{cos}\, ,\\
\sin(\alpha)&=&\frac{r\sin(\beta)}{\sqrt{r^2+d^2+2dr\cos(\beta)}}\label{sin}\, .
\eea
Note that eq.~\eqref{xirelalpha} agrees with the previous calculation of~\cite{francesco, bertacca}.

Expression~\eqref{xirelalpha} is valid in the full-sky. For small angular separations between the bright and faint galaxies it can be further simplified. Indeed in this case $d\ll r$ so that expressions~\eqref{rp} to~\eqref{sin} can be expanded in series of $d/r\ll1$. Moreover, the coefficients $R_1$ to $R_6$ are smooth and slowly varying functions (see appendix~\ref{app:derivationrel}). In the regime where $|r'-r|\ll r$ these functions can be Taylor expanded around $r$
\bea
g(r')&=&g(r+r'-r)\simeq g(r)+g'(r)(r'-r)\nonumber\\
&\simeq& g(r)+rg'(r)\frac{d}{r}\cos(\beta)\, ,\label{evol}
\eea
where here a prime denotes a derivative with respect to $r$.
In this regime, at lowest order in $d/r$, expression~\eqref{xirelalpha} becomes
\begin{align}
\xirel(r, & d, \beta)=\frac{2A}{9\Omega_m^2\pi^2}\frac{\HH}{\HH_0} D_1^2 f \Bigg\{\label{xirelcos}\\
&\nu_1(d) \cdot P_1(\cos\beta)
\Bigg[\left(\frac{\dot{\HH}}{\HH^2}+\frac{2}{r\HH}\right)(\bB-\bF)\nonumber \\
&+\left(\frac{1}{r\HH}-1\right)\Big(5\big(s_\B\bB-s_\F\bF\big)
+3(s_\B-s_\F)f \Big)\Bigg]\nonumber \\
&+ \nu_3(d) \cdot P_3(\cos\beta)\,  2\left(1-\frac{1}{r\HH}\right)(s_\B-s_\F)f\Bigg\}  \, ,\nonumber
\end{align}
where all the functions are evaluated at $r$. Here $D_1$ is the linear growth factor, $P_1$ and $P_3$ are the Legendre polynomial of degree 1 and 3 respectively and
\begin{align}
f&=\frac{d\ln D_1}{d \ln a}\, ,\\
\nu_\ell(d)&=\int\frac{dk}{k}\left(\frac{k}{H_0}\right)^3(k\eta_0)^{n_s-1}T^2(k)j_\ell(kd)\,, \hspace{0.1cm} \ell=1,3\, .\nonumber
\end{align}
We see that the relativistic correlation function is completely anti-symmetric around the bright galaxy: $\xirel(r, d, \beta+\pi)=-\xirel(r, d,\beta)$. It vanishes when $b_\B=b_\F$ and $s_\B=s_\F$. This shows that to measure the relativistic correlation function at lowest order in $d/r$ it is essential to have two distinct populations of galaxies. From eq.~\eqref{xirelcos} we see that if there is no magnification bias, $s_\B=s_\F=0$, then the relativistic term has only a dipolar modulation. With magnification bias however, we have also an octupole modulation.

At higher order in $d/r$, eq.~\eqref{xirelcos} receives two types of corrections. First, the evolution of the density and velocity growth factor generates corrections to the coefficients $R_1$ to $R_6$, that can be calculated using eq.~\eqref{evol}. Second, observations are performed on our past-light cone, and the directions $\bn$ and $\bn'$ are not parallel. In the wide-angle regime, this induces geometrical corrections to the relativistic correlation function that can be calculated by expanding eqs.~\eqref{cos} and~\eqref{sin} at next order in $d/r$. At linear order, $d/r$, these corrections produce a quadrupole, $P_2(\beta)$, and hexadecapole, $P_4(\beta)$, modulation. From this we understand that the contribution of the relativistic terms to the even multipoles is strongly suppressed with respect to the standard density and redshift-space contributions. In addition to the well-known $\HH/k$ suppression, there is an extra suppression due to the factor $d/r$. This is consistent with the analysis of~\cite{seljak} which shows that the relativistic terms have a very negligible effect on the even multipoles. On the other hand, from eq.~\eqref{xirelcos} it is clear that the dipole does not suffer from this extra $d/r$ suppression and is therefore much more promising to measure relativistic corrections.

\subsection{Standard contribution}
\label{sec:standard}

We now perform the same derivation for the standard correlation function $\xist$. As shown in~\cite{szapudi, szapudi2, francesco} in the full-sky $\xist$ reads
\begin{align}
\xist=&\frac{2A}{9\pi^2\Omega_m^2}\Big\{S_1+S_2\cos(2\beta)+S_3\cos(2\alpha)\label{xist}\\
&+S_4\cos(2\alpha)\cos(2\beta)
+S_5\sin(2\alpha)\sin(2\beta) \Big\}\, ,\nonumber
\end{align}
where the coefficients $S_1$ to $S_5$ are given in appendix~\ref{app:coeffst}. They depend on $r,\;r'$ and $d$. Using eqs.~\eqref{rp} to~\eqref{sin}, we can express $\xist$ as a function of $r, d$ and $\beta$. In the small-angle regime, and using the expansion~\eqref{evol} we find that at lowest order in $d/r$, the standard correlation function $\xist$ takes the form
\begin{align}
\xist(r, d, \beta)&= \frac{2AD_1^2}{9\pi^2\Omega_m^2}\Bigg\{\Bigg[\bB\bF+(\bB+\bF)\frac{f}{3}+ \frac{f^2}{5}\Bigg]\mu_0(d)
\nonumber\\
& -\Bigg[(\bB+\bF)\frac{2f}{3} +\frac{4f^2}{7}  \Bigg]\mu_2(d) \cdot P_2(\cos\beta)\nonumber\\
&+\frac{8f^2}{35}\mu_4(d)\cdot P_4(\cos\beta)\Bigg\}\, ,  \label{stmono}
\end{align}
where all the functions are evaluated at $r$ and
\begin{align}
\mu_\ell(d)&=\int\frac{dk}{k}\left(\frac{k}{H_0}\right)^4(k\eta_0)^{n_s-1}T^2(k)j_\ell(kd)\,, \hspace{0.1cm} \ell=0,2,4\,.\nonumber
\end{align}
At lowest order in $d/r$ we see that the standard correlation function is completely symmetric: it contains a monopole, a quadrupole and an hexadecapole. This is simply the well-known Kaiser formula~\cite{kaiserformula, hamilton, cole}. 

Comparing eq.~\eqref{xirelcos} and eq.~\eqref{stmono} we see that at order zero in $d/r$, the odd multipoles are only generated by the relativistic terms, whereas the even multipoles are only generated by the standard terms. The anti-symmetric correlation function provides therefore in principle a clean way to isolate the relativistic terms from the standard ones. This distinction between the standard and relativistic terms based on their symmetry has already been discussed by McDonald in~\cite{mcdonald}. Instead of looking at the correlation function, Ref.~\cite{mcdonald} studies the power spectrum and shows that the relativistic terms contribute to the imaginary part of the power spectrum whereas the standard terms contribute to the real part. 

This distinction (in redshift space or in Fourier space) suffers however from a caveat. Indeed since the amplitude of the standard terms is enhanced by a factor $k/\HH\sim r/d$ with respect to the relativistic terms, it may not be sufficient to keep only the lowest order in $d/r$ in the standard term. Even in the regime where $d/r\ll 1$ it is possible that the next-to-leading order in the standard terms, i.e. the term linear in $d/r$, contaminates the lowest order in the relativistic terms, $(d/r)^0$.

Let us therefore calculate the linear order $d/r$. We are interested in the anti-symmetric correlation function at this order
\be
\xist_\A=\frac{1}{2}\Big[\xist(r,r',d)-\xist(r',r,d) \Big]\, .
\ee
The coefficients $S_1, S_2$ and $S_3$ are not symmetric under the exchange of $r$ and $r'$ due to their dependence on the bias $b_\B$ and $b_\F$. They will therefore contribute to the anti-symmetric part of the correlation function. As an example, let us look at the first term in the coefficient $S_1$, which is of the form:
\be
b_\B(r)b_F(r')D_1(r)D_1(r')\mu_0(d)\, .\nonumber
\ee
The anti-symmetric part of this term is
\begin{align}
&\frac{1}{2}\Big[b_\B(r)b_F(r')-b_\B(r')b_F(r)\Big]D_1(r)D_1(r')\mu_0(d)\nonumber\\
\simeq&\frac{r}{2}\Big[b_\B(r)b'_F(r)-b'_\B(r)b_F(r)\Big]D^2_1(r)\mu_0(d)\frac{d}{r}\cos(\beta)\, ,\nonumber
\end{align}
where in the second line we have used the expansion~\eqref{evol} and kept only the linear order in $d/r$. We see therefore that the difference in the evolution of the bright and faint galaxies generates a dipolar asymmetry in the standard correlation function.

In addition to this kind of asymmetry due to evolution, there is another asymmetry related to the fact that we observe on our past light cone. As a consequence, under the exchange of $r$ and $r'$, $\beta$ does not transform as $\beta+\pi$ but as $\beta+\pi-\theta=\alpha+\pi$, where $\theta$ denotes the angular separation between the two galaxies (see Fig.~\ref{fig:cone}). This generates an additional dipolar modulation in $\xist$. In particular,  at linear order in $d/r$, the term $S_2(r,r')\cos(2\beta)$ in eq.~\eqref{xist} becomes under the exchange of $r$ and $r'$
\be
S_2(r',r)\Big[\cos(2\beta)+4\frac{d}{r}\Big(\cos{\beta}-\cos^3{\beta}\Big)\Big]\, ,\nonumber
\ee
and the term 
\be
S_3(r,r')\cos(2\alpha)\simeq S_3(r,r')\Big[\cos(2\beta)+4\frac{d}{r}\Big(\cos{\beta}-\cos^3{\beta}\Big)\Big]\, ,\nonumber
\ee
becomes under the exchange of $r$ and $r'$
\be
S_3(r',r)\cos(2\beta)\, .\nonumber
\ee
These terms therefore generate an additional asymmetry in the correlation function of the form $4d/r(\cos{\beta}-\cos^3\beta)$. Note that the terms $S_4$ and $S_5$ do not contribute to the anti-symmetric correlation function since they are symmetric $S_4(r,r')=S_4(r',r)$ and $S_5(r,r')=S_5(r',r)$ and their associated angular dependence is also symmetric: $\cos(2\alpha)\cos(2\beta)$ and $\sin(2\alpha)\sin(2\beta)$ do not change under the exchange of $r$ and $r'$. Putting everything together we find that the anti-symmetric correlation function has a total dipolar modulation of the form 
\begin{align}
\xist_{\rm dip}(r, d, \beta)&=\frac{2AD_1^2}{9\pi^2\Omega_m^2}\Bigg\{
-(b_\B-b_\F)\frac{2f}{5}\cdot\mu_2(d)\label{stdip}\\
&+(b_\B-b_\F)\frac{rf'}{6}\Big[\mu_0(d)-\frac{4}{5}\mu_2(d)\Big] \nonumber\\
&-\frac{rf}{6}\big(b'_\B-b'_\F\big)\Big[\mu_0(d)-\frac{4}{5}\mu_2(d)\Big]\nonumber\\
&+\frac{r}{2}\big(b_\B b'_\F-b'_\B b_\F\big)\cdot\mu_0(d)\Bigg\}\cdot\frac{d}{r}\cdot P_1(\cos\beta)\, ,\nonumber
\end{align}
and an octupole modulation of the form
\begin{align}
\label{stoct}
\xist_{\rm oct}(r, d, \beta)&=\frac{2AD_1^2}{9\pi^2\Omega_m^2}\Bigg\{
(b_\B-b_\F)\frac{2f}{5} -(b_\B-b_\F)\frac{rf'}{5}\nonumber\\
&+\big(b'_\B-b'_\F\big)\frac{rf}{5}\Bigg\}\mu_2(d)\cdot\frac{d}{r}\cdot P_3(\cos\beta)\, ,
\end{align}
where all the functions are evaluated at position $r$ and a prime denotes a derivative with respect to $r$. 
Comparing eq.~\eqref{stdip} with the dipolar modulation of the relativistic term, eq.~\eqref{xirelcos}, we see that the standard contribution is suppressed by a factor $d/r$ with respect to the relativistic contribution. As discussed before this reflects the fact that the relativistic term is intrinsically anti-symmetric, due to its direct dependence in the line-of-sight peculiar velocity, whereas the standard term is intrinsically symmetric. The asymmetry of the standard term comes from the evolution of the bias and growth factor (lines 2 to 4 in eq.~\eqref{stdip}) as well as from the wide-angle correction related to the fact that we observe on our past light-cone (first line). As such this dipolar modulation tends to zero with $d/r$.  
On the other hand, the standard term depends on the functions $\mu_\ell(d)$ that are a factor $k/\HH$ larger than the functions $\nu_\ell(d)$ in the relativistic term. Hence eqs.~\eqref{stdip} and~\eqref{stoct} can potentially be of the same order of magnitude as eq.~\eqref{xirelcos}.

\subsection{Lensing contribution}
\label{sec:lensing}

The lensing contribution in eq.~\eqref{xilens} contains three types of term: the correlation between the density of the bright galaxies and the lensing of the faint galaxies $\xilens_{\rm BL}$, the correlation between the density of the faint galaxies and the lensing of the bright galaxies $\xilens_{\rm FL}$, and the lensing-lensing correlation $\xilens_{\rm LL}$. These three terms can be calculated using Limber approximation, in which the contributions of $k_\parallel=\bk\cdot\bn$ to the power spectrum are neglected. In this approximation the correlation between redshift-space distortions (sensitive to $k_\parallel$ only) and lensing vanishes. The first term reads
\begin{align}
\xilens_{\rm BL}\simeq&\frac{2A}{3\Omega_m\pi} b_\B(r)\Big(5s_\F(r')-2\Big)\frac{r(r'-r)}{2r'}\frac{D_1^2(r)}{a(r)}\theta(r'-r)\nonumber\\
&\int_0^\infty dk_\perp \left(\frac{k_\perp}{\HH_0}\right)^2T^2(k_\perp)J_0(k_\perp |\Delta \bx_\perp|)\label{lenslimb} \, ,
\end{align}
where $\theta(x)$ is the Heaviside function: $\theta(x)=1$ if $x>0$ and zero elsewhere.
Eq.~\eqref{lenslimb} describes how the density of the faint galaxies is lensed by the density of the bright galaxies that are in front, i.e. with $r<r'$. The second contribution, $\xilens_{\rm FL}$, describes how the density of the bright galaxies can be lensed by the density of the faint galaxies in front of the bright. This term has exactly the same form as eq.~\eqref{lenslimb}, with $b_\B$ replaced by $b_\F$, $s_\F$ replaced by $s_\B$ and $r$ and $r'$ exchanged. It contributes to the correlation function only if $r>r'$. To that we need to add the lensing-lensing correlation that can also be simplified using Limber approximation.
Putting the three terms together and keeping only the lowest order in $d/r$ we find
\begin{align}
&\xilens=\frac{2A}{3\Omega_m\pi}\cdot\int_0^\infty \frac{dk_\perp}{k_\perp} \left(\frac{k_\perp}{\HH_0}\right)^3T^2(k_\perp)
J_0\big(k_\perp |\Delta x_\perp|\big)\nonumber\\
&\Bigg\{3\Omega_m\HH_0^3(5s_\B-2)(5s_\F-2)\int_0^r d\tilde{r}
 \frac{D_1^2(\tilde{r})}{a^2(\tilde{r})}\frac{(r-\tilde{r})^2\tilde{r}^2}{4 r^2}\nonumber\\
&+\Bigg[\bB(5s_\F-2)\theta\big(\cos\beta\big) -\bF(5s_\B-2)\theta\big(-\cos\beta\big)\Bigg]\nonumber\\
&\hspace{0.5cm}\cdot\frac{D_1^2(r)}{2 a(r)}\HH_0d\cos(\beta)\Bigg\}\, .\label{lenssym}
\end{align}
The multipole expansion of this expression is not trivial, due to the dependence in $\beta$ of the argument of the Bessel function 
$|\Delta x_\perp|=d\sqrt{1-\cos^2(\beta)}$. 
The multipoles can however easily be calculated numerically by projecting eq.~\eqref{lenssym} over the Legendre polynomial
\be
\xilens_\ell=\frac{2\ell+1}{2}\int_{-1}^{1}d\mu\, \xilens(r, d, \mu)P_\ell(\mu)\, , \label{lensmult}
\ee
where $\mu=\cos(\beta)$.
If one considers only one galaxy population, $\bB=\bF$ and $s_\B=s_\F$, the lensing contribution eq.~\eqref{lenssym} is completely symmetric and only the even multipoles are non-zero (see~\cite{lensingHGV} for a detailed study of the lensing anisotropies with one population). However, when $b_\B\neq b_\F$ and $s_\B\neq s_\F$ there is a an anti-symmetric contribution to the correlation function, due to the third line of eq.~\eqref{lenssym}, which reads
\begin{align}
&\xilens_\A=\frac{1}{2}\Big[\xilens(r,r',d)-\xilens(r',r,d)\Big]\label{lensanti}\\
&\simeq \frac{A}{6\pi\Omega_m}\frac{D_1^2(r)}{a(r)}\Big[b_\B(5s_\F-2)-b_\F(5s_\B-2)\Big]\HH_0 d\cos(\beta)\nonumber\\
&\cdot \int_0^\infty \frac{dk_\perp}{k_\perp}\left(\frac{k_\perp}{\HH_0}\right)^3
T^2(k_\perp)J_0\big(k_\perp d\sqrt{1-\cos^2(\beta)}\big)\, .\nonumber
\end{align}
The lensing-lensing correlation does not contribution to this anti-symmetry at lowest order, where $r=r'$, but it would contribute at next-to-leading order. 
The dipolar contribution is then simply
\be
\xilens_{\rm dip}=\frac{3}{2}\int_{-1}^{1}d\mu\,\xilens_\A(r, d, \mu)P_1(\mu)\, .\label{lensdip}
\ee
Comparing eq.~\eqref{lensanti} with the relativistic and standard dipolar contributions~eqs.~\eqref{xirelcos} and~\eqref{stdip}, we see that the lensing contribution is suppressed by a factor $\HH_0 d$ with respect to the relativistic contribution, and by a factor $\HH_0/k\sim \HH_0 d$ with respect to the standard dipole contribution. This suppression is due to the fact that the lensing is sensitive to the projection of the density perpendicular to the line-of-sight. Consequently, the lensing contribution to the dipole will be negligible with respect to both the relativistic and standard contributions. 

\subsection{Alcock-Paczynski contribution}
\label{sec:APcorr}

The Alcock-Paczynski contribution in eq.~\eqref{xiAP} can be rewritten as  
\begin{align}
\xiAP=&\big\langle \Delta_\B(r, \bn)\big(\dd_{r'}-\dd_{\eta'} \big)\Delta_\F(r',\bn') \big\rangle \frac{\partial r'}{\partial \Omega}\cdot\delta\Omega\nonumber\\
+&\big\langle \big(\dd_{r}-\dd_{\eta} \big)\Delta_\B(r,\bn) \Delta_\F(r', \bn')\big\rangle \frac{\partial r}{\partial \Omega}\cdot\delta\Omega\, ,
\label{xiAPcalc}
\end{align}
where $\Delta_\B$ regroups the standard, relativistic and lensing contributions:
\be
\Delta_\B=\Dst_\B+\Drel_\B+\Dlens_\B\, ,
\ee
and similarly for $\Delta_\F$. We can Taylor-expand $r'(z',\Omega)$ around $z$ so that
\begin{align}
\frac{\dd r'(z',\Omega)}{\dd \Omega}\delta\Omega&\simeq\left[\frac{\dd r(z,\Omega)}{\dd \Omega}-\frac{z'-z}{\HH^2(1+z)}\frac{\dd\HH}{\dd\Omega}\right]\delta\Omega\nonumber\\
&=r\left[\frac{\dd\ln r}{\dd\ln\Omega}-\frac{\dd\ln\HH}{\dd\ln\Omega}\,\frac{d}{r}\cos\beta\right]\frac{\delta\Omega}{\Omega}\, .
\label{drexp}
\end{align}

\subsubsection{Radial derivatives}

Let us start by looking at the terms that contains radial derivatives in eq.~\eqref{xiAPcalc}. Using eq.~\eqref{drexp} the anti-symmetric contribution generated by this term is
\begin{align}
\label{xiAPradial}
\xiAP_\A&=\frac{r}{2}\big(\dd_r+\dd_{r'} \big)\Big[\langle \Delta_\B(r,\bn)\Delta_\F(r',\bn')\rangle\\
&-\langle \Delta_\F(r,\bn)\Delta_\B(r',\bn')\rangle\Big]\frac{\dd\ln r}{\dd\ln\Omega}\cdot \frac{\delta\Omega}{\Omega}\nonumber\\
&- \frac{r}{2}\dd_{r'}\Big[\langle \Delta_\B(r,\bn)\Delta_\F(r',\bn')\rangle\nonumber\\
&-\langle \Delta_\F(r,\bn)\Delta_\B(r',\bn')\rangle\Big]\frac{d}{r}\cos\beta\,\frac{\dd\ln\HH}{\dd\ln\Omega}\cdot \frac{\delta\Omega}{\Omega}\,  .\nonumber
\end{align}
The contribution from the first two lines brings in a factor $kr (i\hk\cdot\bn'-i\hk\cdot\bn)$. In the flat sky approximation, $\bn'=\bn$, and this term vanishes. In the full sky however, there is a contribution proportional to $kr (\cos\beta-\cos\alpha)\simeq kr(\cos^2\beta-1)d/r$. We see that the enhancement  brought by the radial derivative ($\sim kr\gg 1$) is counterbalanced by the wide-angle correction ($\sim d/r\ll1$). As a consequence this contribution is roughly of the order of
\begin{align}
&\frac{1}{2}\Big[\langle \Delta_\B(r,\bn)\Delta_\F(r',\bn')\rangle
-\langle \Delta_\F(r,\bn)\Delta_\B(r',\bn')\rangle\Big]\nonumber\\
&\cdot(\cos^2\beta-1)\,
\frac{\dd\ln r}{\dd\ln\Omega}\cdot \frac{\delta\Omega}{\Omega}\, .\nonumber
\end{align}
It is only generated by the anti-symmetric terms in the correlation function, and it is further suppressed by $\delta\Omega/\Omega\lsim 0.01$.

The contribution from the two last lines in eq.~\eqref{xiAPradial} is of a similar order. The radial derivative generates a factor $ikr\hk\cdot\bn'$ which, combined with the other terms, gives a contribution proportional to $kr \cos\beta\cos\alpha \,d/r\simeq kr \cos^2\beta \,d/r$. As previously, the factor $d/r$ counterbalances the enhancement $kr$, so that the contribution is of the order
\begin{align}
&\frac{1}{2}\Big[\langle \Delta_\B(r,\bn)\Delta_\F(r',\bn')\rangle
-\langle \Delta_\F(r,\bn)\Delta_\B(r',\bn')\rangle\Big]\nonumber\\
&\cdot\cos^2\beta\,
\frac{\dd\ln \HH}{\dd\ln\Omega}\cdot \frac{\delta\Omega}{\Omega}\, .\nonumber
\end{align}
This term as well is only generated by the anti-symmetric terms in the correlation function, and it is further suppressed by $\delta\Omega/\Omega\lsim 0.01$.

These anti-symmetric contributions are therefore negligible with respect to the relativistic and evolutionary standard terms. In appendix~\ref{app:AP} we give an explicit calculation of one of these contributions. 

\subsubsection{Time derivatives}

The anti-symmetric contributions with time derivatives in eq.~\eqref{xiAPcalc} read
\begin{align}
\label{xiAPtime}
\xiAP_\A&=-\frac{1}{2}\big(\dd_\eta+\dd_{\eta'} \big)\Big[\langle \Delta_\B(r,\bn)\Delta_\F(r',\bn')\rangle\\
&-\langle \Delta_\F(r,\bn)\Delta_\B(r',\bn')\rangle\Big]\frac{\dd\ln r}{\dd\ln\Omega}\cdot \frac{\delta\Omega}{\Omega}\nonumber\\
&+\frac{1}{2}\dd_{\eta'}\Big[\langle \Delta_\B(r,\bn)\Delta_\F(r',\bn')\rangle\nonumber\\
&-\langle \Delta_\F(r,\bn)\Delta_\B(r',\bn')\rangle\Big]\frac{d}{r}\,\cos\beta\,\frac{\dd\ln\HH}{\dd\ln\Omega}\cdot\frac{\delta\Omega}{\Omega}\,  .\nonumber
\end{align}
Contrary to the radial derivatives, the time derivatives $\dd_\eta$ and $\dd_{\eta'}$ do not bring any additional $k$-factor. Using that $\dd_\eta\Delta$ is roughly of the same order of magnitude as $\Delta$ and that the time derivatives do not change the symmetry of the various contributions, we see immediately that the contribution from the first two lines is only generated by the anti-symmetric correlation function and that it is further suppressed by $\delta\Omega/\Omega$. The contribution from the two last lines is manifestly anti-symmetric due to the $\cos\beta$, and it is also suppressed with respect to the standard terms that have a similar factor $d/r$ but not the extra factor $\delta\Omega/\Omega$. In appendix~\ref{app:AP} we give an explicit calculation of one of the time derivative contributions. 

In summary, this calculation shows that the Alcock-Paczynski effect does not generate any new anti-symmetric terms in the correlation function. It only induces corrections to the anti-symmetric contributions that already exist (namely those created by the relativistic terms and the evolution of the standard terms). Since the  Alcock-Paczynski corrections are suppressed by a factor $\delta\Omega/\Omega\lsim 0.01$, we can safely neglect them in our analysis.

\begin{figure}[t]
\centerline{\epsfig{figure=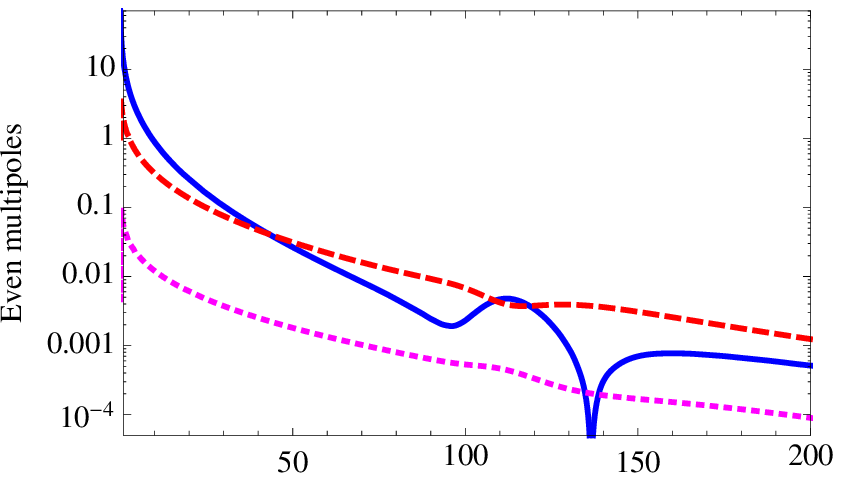,height=5.1cm}}\vspace{-0.3cm}
\centerline{\epsfig{figure=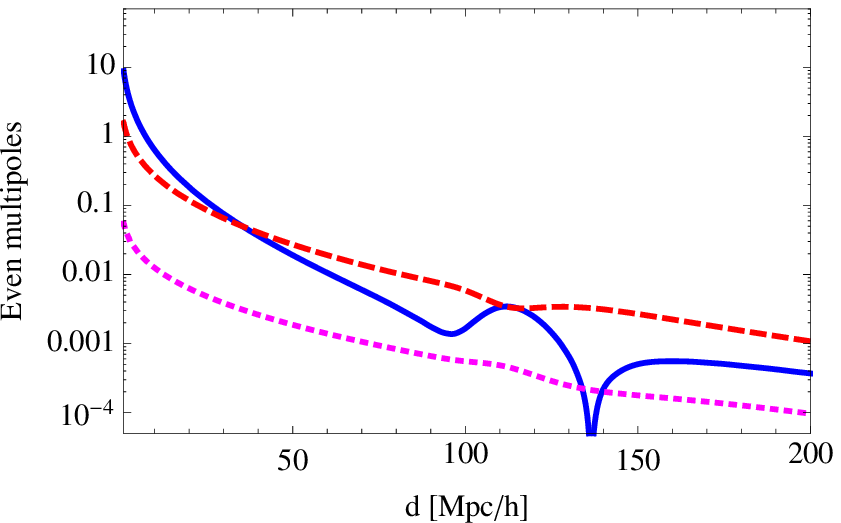,height=5.1cm}}
\caption{ \label{fig:mult} The monopole (blue solid line), quadrupole (red dashed line) and hexadecapole (magenta dotted line) plotted as a function of $d$. The top panel is for $z_\B=0.25$ and the bottom panel for $z_\B=1$. In both cases, the monopole is positive at small scales and negative at large scales. The quadrupole is always negative, whereas the hexadecapole is always positive.}
\end{figure}

\begin{figure}[!h]
\centerline{\epsfig{figure=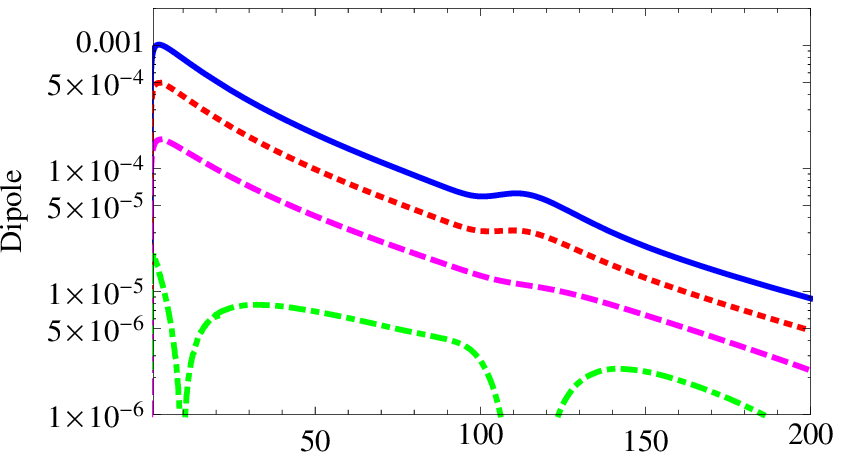,height=5.1cm}}\vspace{-0.3cm}
\centerline{\epsfig{figure=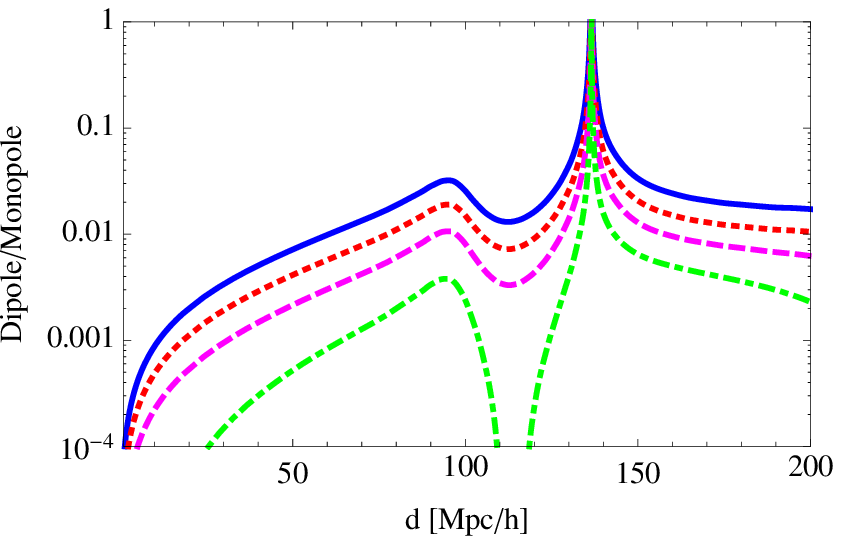,height=5.1cm}}
\caption{ \label{fig:diptot} Top panel: the total dipole as a function of $d$ at $z_\B=0.25$ (blue solid line), $z_\B=0.5$ (red dotted line), $z_\B=1$ (magenta dashed line) and $z_\B=2$ (green dash-dotted line). The dipole is positive, except at $z_\B=2$ where it changes sign.
Bottom panel: the ratio of the total dipole over the monopole for the same four values of $z_\B$.}
\end{figure}

\begin{figure}[!h]
\centerline{\epsfig{figure=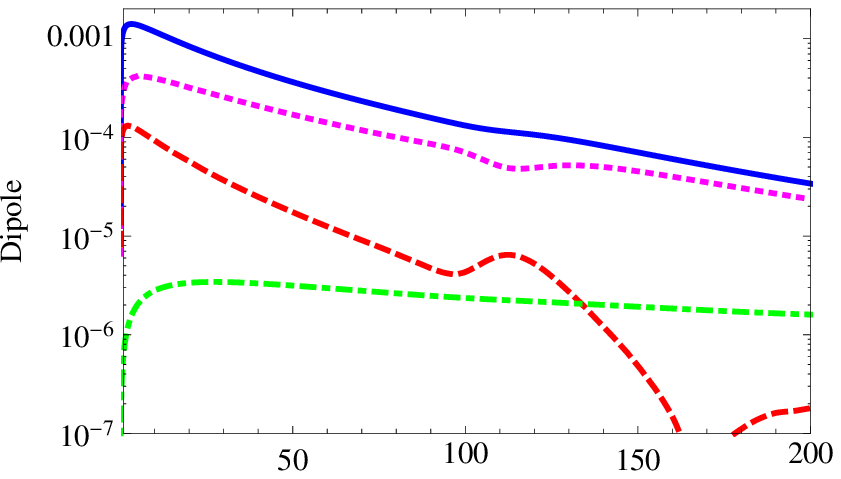,height=5.1cm}}\vspace{-0.35cm}
\centerline{\epsfig{figure=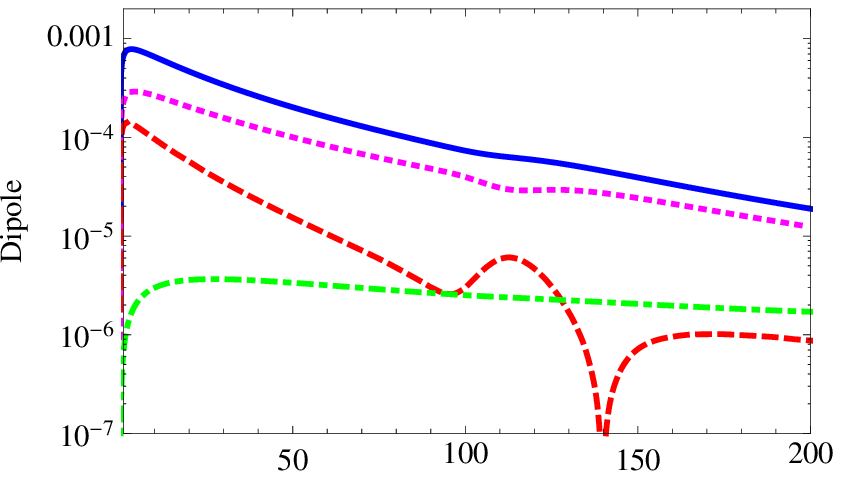,height=5.1cm}}\vspace{-0.35cm}
\centerline{\epsfig{figure=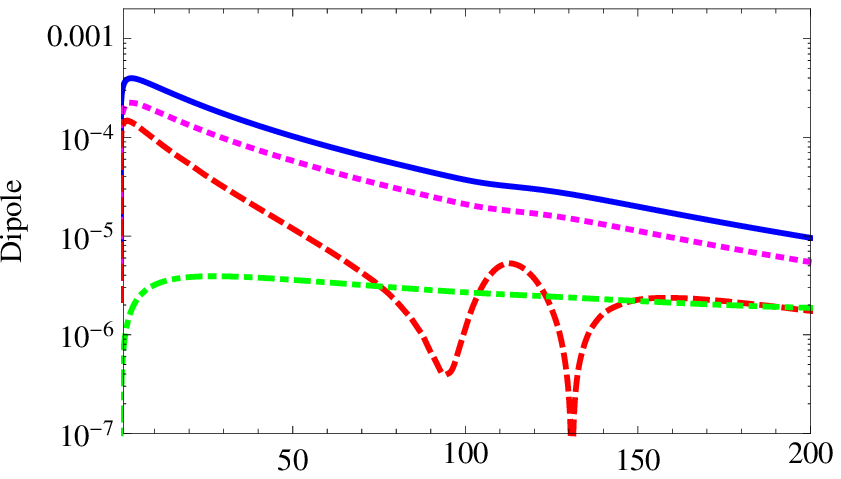,height=5.1cm}}\vspace{-0.35cm}
\centerline{\epsfig{figure=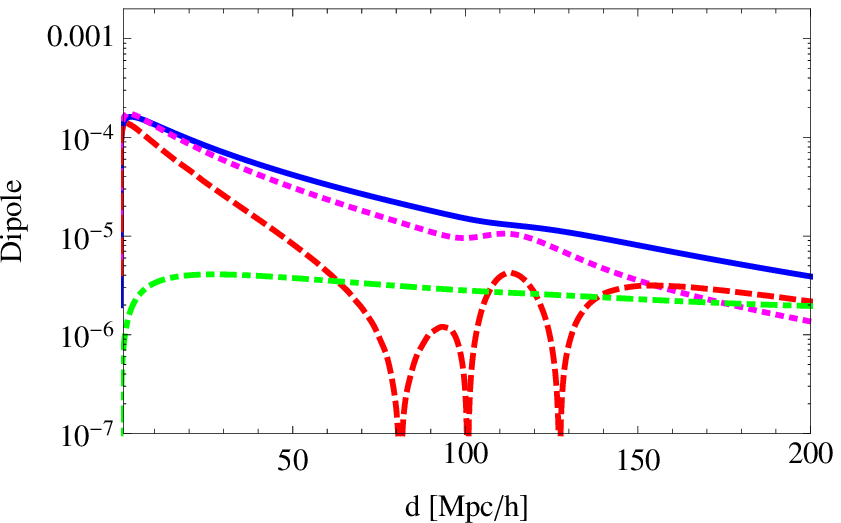,height=5.1cm}}
\caption{ \label{fig:dip} The various dipoles as a function of $d$: the relativistic contribution (blue solid), the standard contribution (magenta dotted), and the lensing contribution (green dash-dotted). The red dashed line is the standard contribution corrected for the wide-angle effect as described in section~\ref{sec:correction}. The four panels are for redshift (from top to bottom) $z_\B=0.25$, $z_\B=0.5$, $z_\B=1$ and $z_\B=2$. The relativistic dipole is positive and the standard and lensing dipoles are negative.
}
\end{figure}

\begin{figure}[!h]
\centerline{\epsfig{figure=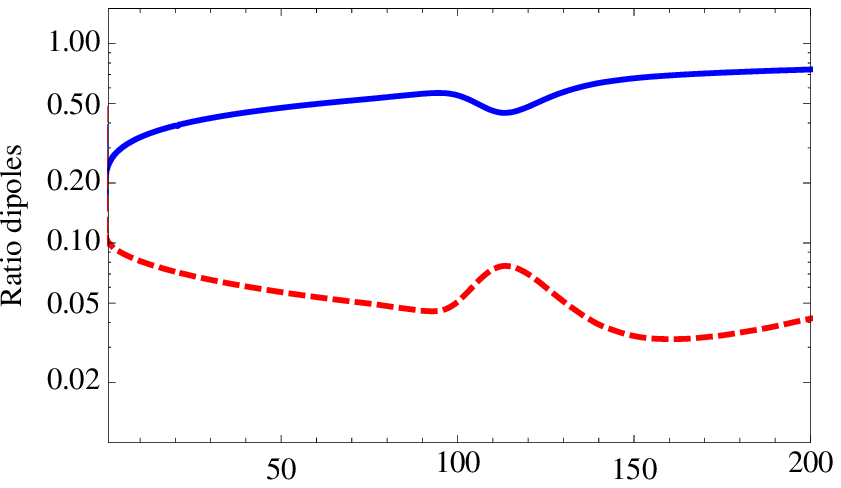,height=5.1cm}}\vspace{-0.35cm}
\centerline{\epsfig{figure=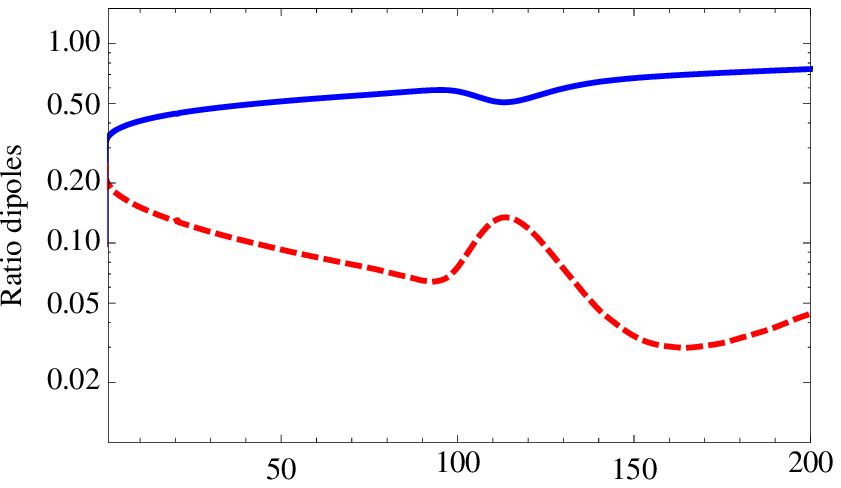,height=5.1cm}}\vspace{-0.35cm}
\centerline{\epsfig{figure=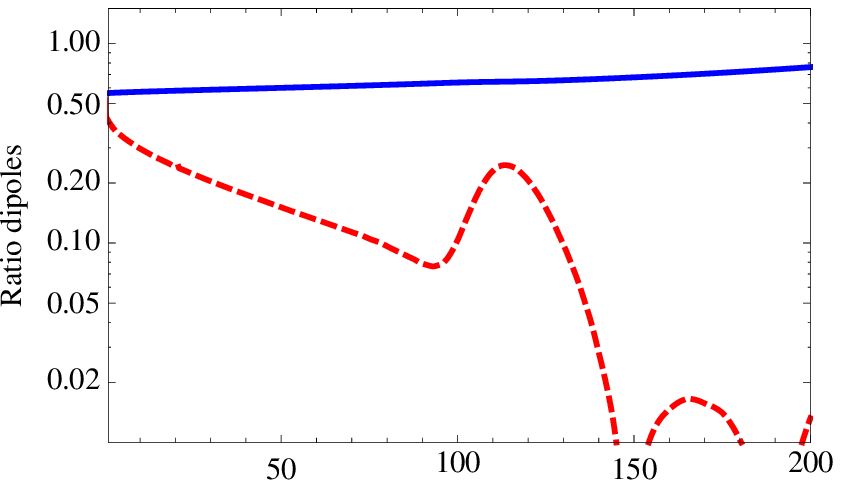,height=5.1cm}}\vspace{-0.35cm}
\centerline{\epsfig{figure=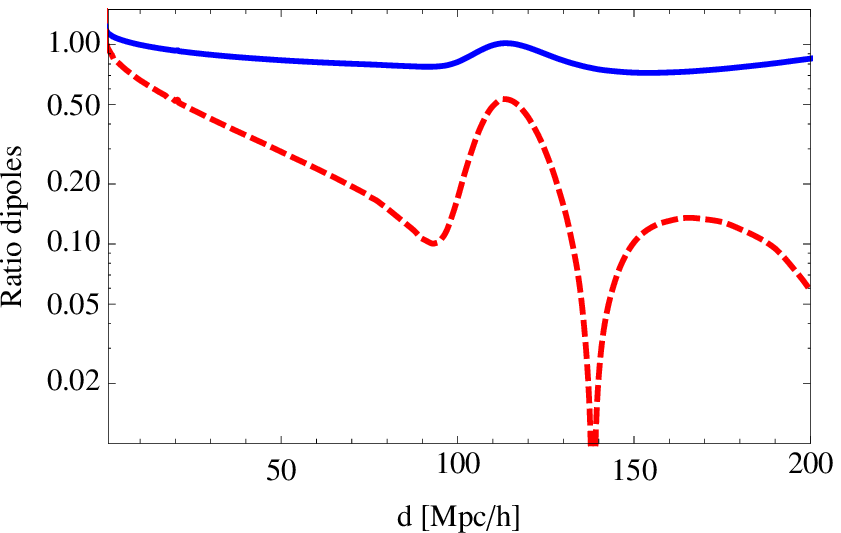,height=5.1cm}}
\caption{ \label{fig:dipratio} The blue solid line shows the ratio between the standard dipole + lensing dipole and the relativistic dipole $(\xist_{\rm dip}+\xilens_{\rm dip})/\xirel_{\rm dip}$, plotted as a function of $d$. The red dashed line shows the ratio between the corrected standard dipole (described in section~\ref{sec:correction}) + lensing dipole, and the relativistic dipole as a function of $d$. The four panels are for redshift (from top to bottom) $z_\B=0.25$, $z_\B=0.5$, $z_\B=1$ and $z_\B=2$.}
\end{figure}

\subsection{Extinction}

Before computing the amplitude of the different terms, let us mention one additional effect that may generate anti-symmetries in the correlation function: dust absorption~\footnote{We thank Jeremiah Ostriker and the anonymous referee for bringing this important point to our attention.}. The existence of cosmic dust has been detected recently in ~\cite{dust}. Measurements of the cross-correlation of the brightness of high redshift quasars with the clustering of foreground galaxies have indicated the presence of dust well beyond the star forming region of the galaxies. In~\cite{dust_anis}, it has been shown that this extended dust, which is correlated with galactic haloes and the large-scale structure, generates anisotropies in the two-point correlation function. Here we also expect that it would generate anti-symmetries in the cross-correlation between faint and bright galaxies. A faint galaxy behind a bright one will indeed be obscured by the dust correlated with the bright galaxy, whereas a faint galaxy in front will not. If the bright and faint galaxies are associated with the same amount of dust, this effect will cancel out in average. But if the dust depends on the galaxy population (which is expected) then it will generate anti-symmetries in the cross-correlation function. We defer the calculation of the this effect, which requires a careful modelling of the dust distribution, to a subsequent paper. Note however that it should be possible to separate this effect from the relativistic term by using the fact that dust extinction depends on the colour of galaxies whereas the relativistic contribution does not.

\section{Results}

\label{sec:results}

\begin{figure}[!t]
\centerline{\epsfig{figure=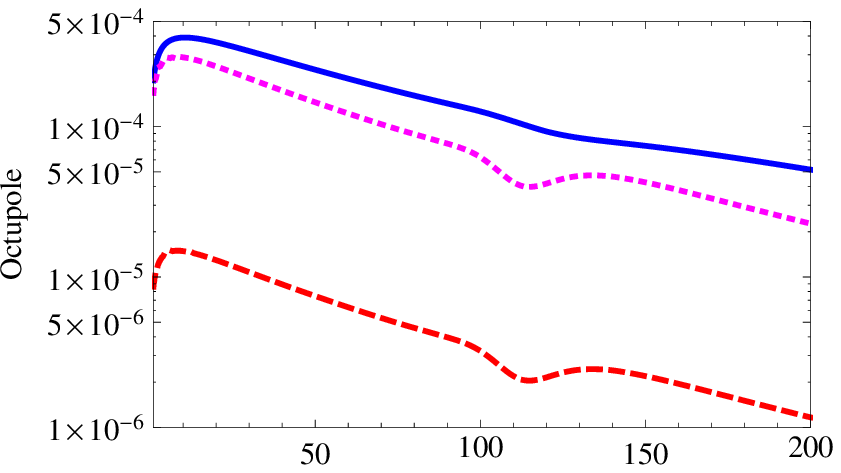,height=5.1cm}}\vspace{-0.35cm}
\centerline{\epsfig{figure=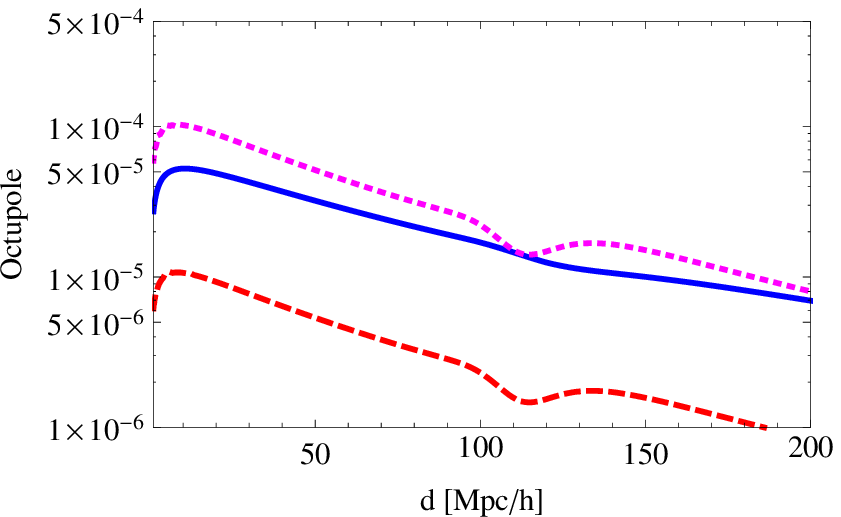,height=5.1cm}}\vspace{-0.35cm}
\caption{ \label{fig:oct} The various octupoles as a function of $d$: the relativistic contribution (blue solid) and the standard contribution (magenta dotted). The red dashed line is the standard contribution corrected for the wide-angle effect as described in section~\ref{sec:correction}. The relativistic octupole has been calculated for $s_\B-s_\F=1$. The top panel is at redshift $z_\B=0.25$ and the bottom panel at $z_\B=1$. The relativistic octupole is negative whereas the standard octupole is positive.
}
\end{figure}

We now compare the relativistic, standard and lensing contributions in a flat $\Lambda$CDM universe. We choose $n_s=1$, $\Omega_m=0.24$, $h=0.73$ and $\sigma_8=0.75$ so that the primordial amplitude of the power spectrum is $A=1.8\times10^{-8}$. We use CAMB~\cite{camb} to compute the transfer function $T(k)$. We ignore magnification bias (except in Fig.~\ref{fig:oct}) so that $s_\B=s_\F=0$ and we assume that the bias of the bright and faint galaxies evolves as~\cite{biasevol}
\begin{align}
b_\B(z)&=1+(b_\B^i-1)\frac{D_1(z_i)}{D_1(z)}\, ,\\
b_\F(z)&=1+(b_\F^i-1)\frac{D_1(z_i)}{D_1(z)}\, ,
\end{align}
where $b_\B^i$ and $b_\F^i$ are the initial values of the bias at redshift $z_i\simeq3$. We choose $b_\B^i$ and $b_\F^i$ such that at $z=0.5$ $b_\B=2$ and $b_\F=1.5$. With this we have that at $z=0.25$: $b_\B=1.9$ and $b_\F=1.45$, at $z=1$: $b_\B=2.25$ and $b_\F=1.62$ and at $z=2$: $b_\B=2.8$ and $b_\F=1.9$.

\subsection{Comparison of the multipoles}

In Fig.~\ref{fig:mult} we plot the amplitude of the monopole, the quadrupole and the hexadecapole, calculated in eq.~\eqref{stmono} and~\eqref{lensmult}. The redshift of the bright galaxy is kept fixed and the multipoles are plotted as a function of the comoving separation $d$. The top panel is for $z_\B=0.25$ and the bottom panel for $z_\B=1$.  We see that at small scales the monopole dominates but at large scales $d\gsim 50\,  h^{-1}$Mpc the quadrupole becomes important. The hexadecapole remains a factor 10 times smaller than the quadrupole at all scales. The lensing contribution to the monopole and quadrupole is smaller than a percent at all scales. The contribution to the hexadecapole is however more important: at  $d\gsim 100\,  h^{-1}$Mpc the lensing contributes to 5\%, whereas at $d\gsim 200\,  h^{-1}$Mpc it reaches 35\%. In all cases the dominant lensing contribution comes from the cross-correlation between the density and lensing. The lensing-lensing correlation is always at least 3 orders of magnitude smaller. 

In Fig.~\ref{fig:diptot} (top panel) we plot the amplitude of the total dipole, i.e. the sum of the relativistic contribution, eq.~ \eqref{xirelcos}, the standard contribution, eq.~\eqref{stdip} and the lensing contribution, eq.~\eqref{lensdip}. The different curves are for different redshift values of the bright galaxy, from $z_\B=0.25$ (top) to $z_\B=2$ (bottom). We see that the dipole decreases with $d$ and with redshift. The bottom panel of Fig.~\ref{fig:diptot} shows the ratio of the dipole over the monopole for the same four values of redshift. The dipole decreases less quickly than the monopole with $d$ so that the ratio becomes larger at large scales. This is due to the fact that the dipole contains relativistic effects that are suppressed by a factor $\HH/k$ with respect to the standard terms. At small $d$, both the monopole and the dipole are sensitive to small wavelengths $k\gg \HH$ and the dipole is therefore strongly suppressed with respect to the monopole. At large $d$ however, the long wavelengths $k\sim \HH$ start to contribute so that the suppression becomes less and less effective.

In Fig.~\ref{fig:dip} we compare the amplitude of the different dipolar contributions: the relativistic term (blue solid line), standard term (magenta dotted line) and lensing term (green dash-dotted line). The meaning of the red dashed line will be explained in section~\ref{sec:correction}. The four panels correspond to four values of the redshift of the bright galaxies $z_\B=0.25,\, 0.5,\, 1$ and 2. 
We see that the relativistic dipole is always positive; the correlation function is therefore stronger for faint galaxies behind the bright galaxy than for faint galaxies in front of the bright galaxy. From eq.~\eqref{xirelcos}, we see that the sign of the relativistic dipole is governed by the bias difference $b_\B-b_\F>0$ (neglecting magnification bias, $s_\B=s_\F=0$). As discussed in more detail in appendix~\ref{app:derivationrel}, this reflects the fact that the dominant contribution to the cross-correlation function is due to the gravitational redshift of the faint galaxies weighted by the density of the bright galaxies, $\delta_\B=b_\B\cdot\delta$.

Interestingly, we see in Fig.~\ref{fig:dip} that the relativistic contribution dominates over the other terms at all redshifts and scales. This shows that the dipole provides a powerful way to measure relativistic corrections in large-scale structure. Comparing the different panels we see that the relative importance of the standard contribution versus the relativistic contribution increases with redshift. The lensing contribution on the other hand remains always subdominant. 

In Fig.~\ref{fig:dipratio} we show the ratio between the standard dipole~+ lensing dipole, and the relativistic dipole (blue line): $(\xist_{\rm dip}+\xilens_{\rm dip})/\xirel_{\rm dip}$. The meaning of the red dashed line will be explained in section~\ref{sec:correction}. We see that at small redshift $z_\B=0.25$ the standard and lensing contributions contaminate the measurement of the relativistic term by~$47$\% at $d=50\, h^{-1}$Mpc. This contamination increases with redshift and reaches $83$\% at $z_\B=2$. 
Below we propose a method to suppress this contamination and disentangle the relativistic terms from the standard terms.

But before doing so let us first look at the octupole modulation, plotted in Fig.~\ref{fig:oct}. If there is no magnification bias, the octupole only receives a contribution from the standard term. With magnification bias however, the relativistic term also contributes to the octupole and is proportional to the slope difference $s_\B-s_\F$.  This difference depends on the characteristics of the survey. In Fig.~\ref{fig:oct} we choose for illustration a difference $s_\B-s_\F=1$. We compare the relativistic octupole (blue solid line) and the standard octupole (magenta dotted line). The meaning of the red dashed line will be explained in section~\ref{sec:correction}. The top panel is for redshift $z_\B=0.25$ and the bottom panel for redshift $z_\B=1$. We see that the relativistic term varies strongly with redshift and becomes smaller than the standard term at $z_\B=1$. This is due to the pre factor $\left(1-\frac{1}{r\HH}\right)$ which decreases with redshift and passes through zero around $z=1.7$. The standard octupole is very similar to the standard dipole: it is dominated by the wide-angle effect which is the same as the one in the dipole, but with opposite sign.

\subsection{Isolating the relativistic contribution}
\label{sec:correction}

Let us now discuss how to reduce the contamination from the standard
term (through wide-angle and evolution effects) in the dipole and the octupole.

\subsubsection{Removing the wide-angle term}

\begin{figure}[t]
\centerline{\epsfig{figure=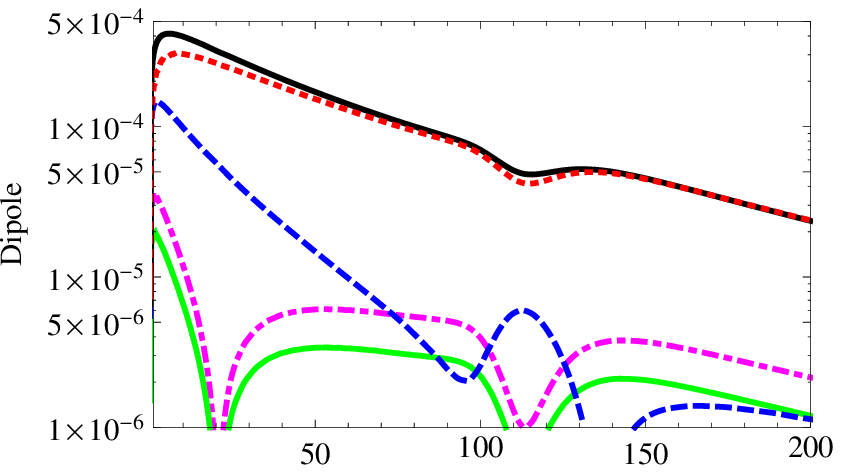,height=5.1cm}}\vspace{-0.3cm}
\centerline{\epsfig{figure=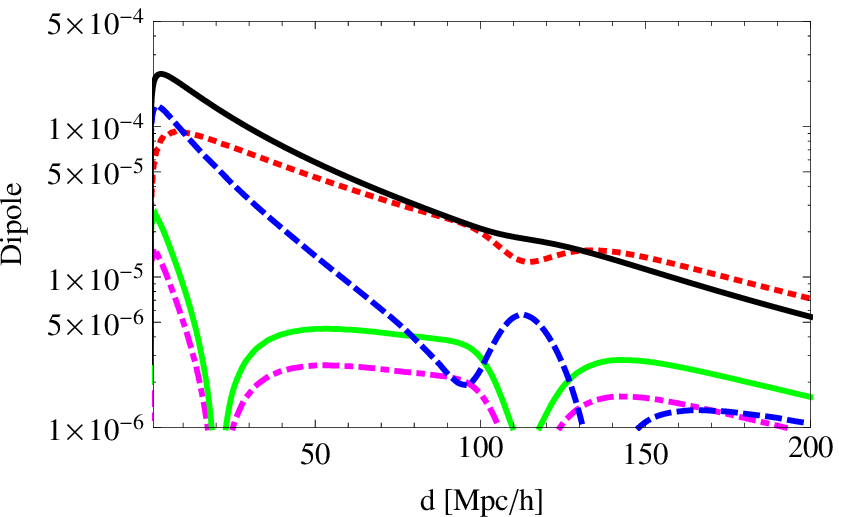,height=5.1cm}}
\caption{ \label{fig:dipstand} The various contributions of eq.~\eqref{stdip} to the standard dipole: the first line 'wide-angle' effect  (red dotted), the second line (magenta dash-dotted), the third line (green solid), the fourth line (blue dashed) and the total (black solid). 
The top panel is at $z_\B=0.25$ and the bottom panel at $z_\B=1$.}
\end{figure}

In Fig.~\ref{fig:dipstand} we compare the various contributions to the standard dipole, calculated in eq.~\eqref{stdip}: the black solid line is the total standard dipole, the red dotted line is the contribution from the first line of eq.~\eqref{stdip}, the magenta dash-dotted line is the contribution from the second line, the green solid line is the contribution from the third line and the dashed blue line the contribution from the last line. We see that the contributions from the first line and the last line dominate at most scales. The term in the first line is a {\it wide-angle} effect, due to the fact that we observe on our past light-cone. The term in the last line on the other hand reflects the fact that the density of the bright and faint galaxies does not evolve in exactly the same way due to the different bias evolution.

We can use the specific form of these terms to remove them from the dipole and octupole modulation and isolate the relativistic terms. From eq.~\eqref{stmono} we see that the quadrupole modulation is very similar to the wide-angle dipolar and octupolar contamination. The quadrupole contributes not only to the cross-correlation between the bright and faint galaxies as in eq.~\eqref{stmono}, but also to the auto-correlation of the bright galaxies and the auto-correlation of the faint galaxies. The difference between these two auto-correlations reads
\begin{align}
&\xist_{\B\, {\rm quad}}(r, d, \beta)-\xist_{\F\, {\rm quad}}(r, d, \beta)=\label{diffquad}\\
&-\frac{2A}{9\pi^2\Omega_m^2}\frac{4}{3}(b_\B-b_\F)D_1^2f\,\mu_2(d)\cdot P_2(\cos\beta)\, .\nonumber
\end{align}
Comparing eq.~\eqref{diffquad} with the first term in eq.~\eqref{stdip} we see that they differ only by a factor $3d/10r$. Hence, we can correct for the wide-angle effect by measuring the quadrupole of the bright and faint populations, taking their difference and multiplying the result by $3d/10r$. In Fig.~\ref{fig:dip} the red dashed line represents the standard dipole after having corrected for the wide-angle effect. This corrected standard dipole is significantly smaller than the relativistic dipole at all redshifts and scales.
In Fig.~\ref{fig:dipratio} the ratio of the corrected standard + lensing dipoles over the relativistic dipole is plotted in dashed red. With this correction, at $d=50\, h^{-1}$Mpc the contamination amounts to only 5\% at redshift $z_\B=0.25$, 9\% at $z_\B=0.5$, 15\% at $z_\B=1$ and 29\% at $z_\B=2$. Hence this correction greatly improves the precision with which one can extract the relativistic contribution from the dipolar modulation. Note that this correction is completely model independent. It does not require any modelling of the density evolution or the bias evolution. It only necessitates a measurement of the quadrupole of the bright and faint populations separately. The exact same correction can be applied to the octupole (with opposite sign) since it contains the same wide-angle term, first term in eq.~\eqref{stoct}. In Fig.~\ref{fig:oct} the red dashed line represents the standard octupole after having corrected for the wide-angle effect. This correction is even more effective for the octupole than for the dipole since the other terms in the octupole are strongly subdominant with respect to the wide-angle effect.

\subsubsection{Removing the evolution effects}

To go further and correct for the other important term in the dipole , i.e. the evolution term in the last line of eq.~\eqref{stdip} is more involved.  The shape of this term is given by $\mu_0(d)$ and is therefore the same as the shape of the monopole term in eq.~\eqref{stmono}. The amplitude however is more complicated to model since it is given by $\big(b_\B b'_\F-b'_\B b_\F\big)D_1^2$ and requires a knowledge of the bias evolution. 

One possibility that can help us determine the bias evolution is to look at asymmetries in the auto-correlation function of the bright and the faint population separately. We argued in section~\ref{sec:intro} that asymmetric correlation functions
can only be found by cross-correlating two populations of galaxies. The validity of this statement depends however on what we call an asymmetry and how the measurement is done. An anti-symmetry under the exchange of the position of two galaxies in the pair can obviously only exist if one cross-correlates two populations. However, an anti-symmetry around a fixed galaxy can exist even in the case where one has only a single galaxy population. 
This asymmetry was first discussed by~\cite{roy}. 

Suppose we select galaxies situated at a redshift $z_*$ (corresponding to a
radial coordinate $r_*$ on our past light-cone). Let us call these
galaxies the {\it central} galaxies. We then correlate the central
galaxies with galaxies behind them (i.e. at a higher redshift) with $r=r_*+\Delta r$, and with
galaxies in front of them (i.e. at a lower redshift) with $r=r_*-\Delta r$.
It is important that in this process, $z_*$ for the central galaxies
is held fixed. The correlation
function $\xi\big(r_*, r_*+\Delta
r,|\Delta \bx_\perp|\big)$ needs not equal
$\xi\big(r_*, r_*-\Delta r, |\Delta
\bx_\perp|\big)$, for some transverse separation $|\Delta \bx_\perp|$, and radial separation $\Delta r$.
Their difference arises entirely from the evolution terms,
and not from the relativistic terms, that cancel out for one population of galaxies.
This shows that the asymmetry around the central galaxies can be used to
isolate evolution effects. Let us emphasize that to do so, one
needs to be careful in the averaging procedure. It is essential to fix
the position of the central galaxies and only average over the {\it other}
galaxies. If one also averages over the 
redshift of the central galaxies within the same volume as the
other galaxies, the evolution asymmetry is washed out,
for in that case, all galaxies are treated on equal footing and there
cannot be any asymmetry.

Denoting by $r_*$ the fixed position of the central galaxies, one finds a dipolar modulation of the form:
\begin{align}
&\xi^{\rm 1 pop.}_{\rm dip}=\frac{2AD_1}{9\pi^2\Omega_m^2}\Bigg\{-\frac{4}{5} b D_1 f\mu_2(d)\\
&+D_1f^2  \left[\frac{8}{25}\mu_0(d)-\frac{4}{7}\mu_2(d)+\frac{54}{175}\mu_4(d)\right]\nonumber\\
&+r\Big[\left(b+\frac{f}{3}\right)(bD_1)'+\left(\frac{b}{3}+\frac{f}{5}\right)(D_1f)' \Big]\mu_0(d)\nonumber\\
&-r\left[\left(\frac{2b}{3}+\frac{4f}{7}\right)(fD_1)'+\frac{2f}{3}(bD_1)'\right]\mu_2(d)\Bigg\}\frac{d}{r_*}P_1(\cos\beta)\nonumber
\end{align}
where $b$ denotes the bias of the galaxy population under
consideration. 
From this dipole -- measured by averaging around a fixed
redshift where the central galaxies are located -- we can
learn something about the bias evolution of each population
separately. This knowledge can then be
used in principle to eliminate the contamination to
the cross-correlation dipole from evolution effects,
though probably in a model-dependent way.

A more ambitious approach would be to
make use of the fact that the evolution effects contribute
to both the symmetric and anti-symmetric parts of the
cross-correlation function, while the relativistic effects
contribute only to the latter. One of the distinguishing
features of the evolution effects is that they break
translational invariance i.e. it matters how far away
the galaxies are from us (and therefore how long ago
they emitted the photons that reach us). This can be used
to isolate the evolution contributions to the symmetric
part of the cross-correlation function, which can then be
used to clean out their contributions to the anti-symmetric
part. Details will be presented in a separate paper.

\subsection{Validity of our approximations}

\begin{figure}[t]
\centerline{\epsfig{figure=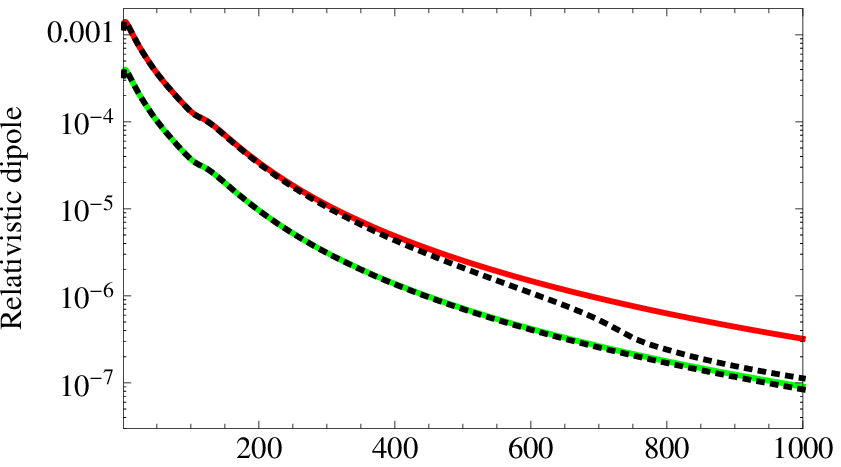,height=5.1cm}}\vspace{-0.3cm}
\centerline{\epsfig{figure=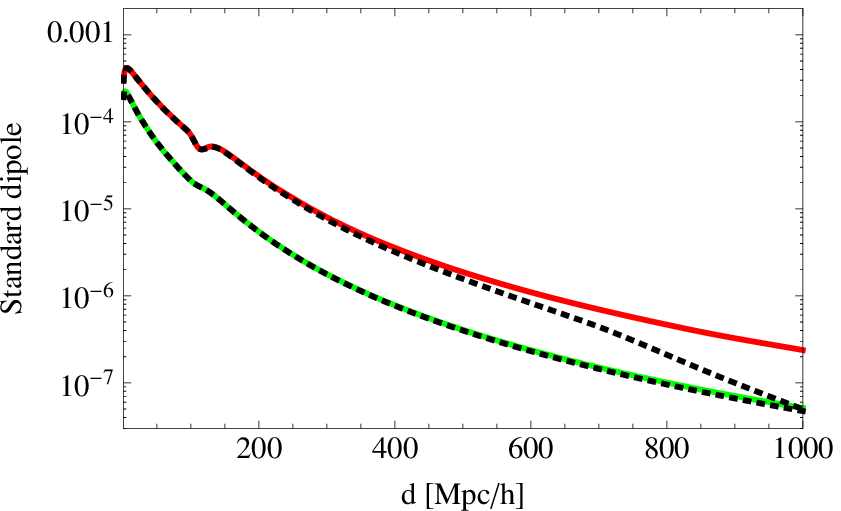,height=5.1cm}}
\caption{ \label{fig:diff} Comparison of our approximations eqs.~\eqref{xirelcos} and~\eqref{stdip} with the full-sky expressions. The top panel shows the relativistic dipole at $z=0.25$ (top curves, red and black) and $z=1$ (bottom curves, green and black). The solid lines are the approximations and the dotted lines the full-sky expressions. The bottom panel shows the same quantities but for the standard dipole.}
\end{figure}

To finish this section, let us discuss the validity of the approximations used in the calculation of the multipoles, in eqs.~\eqref{xirelcos},~\eqref{stmono},~\eqref{stdip} and~\eqref{stoct}. In these expressions we have performed an expansion in $d/r\ll1$ and kept only the lowest order contributions. For the standard terms we have found that the lowest order contribution to the monopole, quadrupole and hexadecapole is $(d/r)^0$, whereas the lowest order contribution to the dipole and octupole is $d/r$. On the other hand for the relativistic terms, we have found that the lowest order contribution to the dipole and octupole is $(d/r)^0$, and the lowest order contribution to the quadrupole and hexadecapole is $d/r$ (the contribution to the monopole is even higher). At small separation, $d\ll r$, these approximations are expected to be accurate. However, at large separation $d\sim r$, wide-angle corrections may become relevant. 

To quantify the importance of these corrections, we can simply use the full-sky expressions, eqs.~\eqref{xirelalpha} and~\eqref{xist}, which can be expressed as functions of $d, r$ and $\beta$ using eqs.~\eqref{rp},~\eqref{cos} and~\eqref{sin}. The even multipoles can be computed numerically through
\be
\xi_\ell=\frac{2\ell+1}{2}\int_{-1}^{1}d\mu\, \xi(r, d, \mu)P_\ell(\mu)\, . \label{multnum}
\ee
At $z=1$, we find that the fractional monopole and quadrupole difference between our approximation~\eqref{stmono} and the full-sky expression is smaller than a percent up to 200$\,  h^{-1}$Mpc. At $z=0.25$, this difference is smaller than a percent up to 100$\,  h^{-1}$Mpc and reaches 7\% at 200$\,  h^{-1}$Mpc. The hexadecapole on the other hand receives larger corrections at wide angle: at $z=1$ the fractional difference is of~6\% at 200$\,  h^{-1}$Mpc, whereas at $z=0.25$ it is already of 20\% at~100$\,  h^{-1}$Mpc. This suggests that at small redshift, the full-sky expression for the hexadecapole should be used for scales larger than 100$\,  h^{-1}$Mpc (see~\cite{rsd1, rsd2, raccanelli} for a detailed study of the full-sky effects on the even multipoles). Note however that here we are comparing the approximation and full-sky expression at fixed redshift. As shown in~\cite{seljak}, averaging over wide redshift bins tends to decrease the difference between small-angle approximations and full-sky expressions since for a given separation $d$, there are more pairs of galaxies at high redshift, where the small-angle approximation is better, than at low redshift.     

To compute the relativistic and standard dipoles in the full sky, we first need to anti-symmetrize eqs.~\eqref{xirelalpha} and~\eqref{xist}, using that under the exchange of $r$ and $r'$, $\beta$ becomes $\alpha+\pi$ and $\alpha$ becomes $\beta+\pi$. We can then compute the dipole with
\be
\xi_1=\frac{3}{2}\int_{-1}^{1}d\mu\, \xi_\A(r, d, \mu)P_1(\mu)\, . \label{dipnum}
\ee
In Fig.~\ref{fig:diff} we plot the relativistic dipole (top panel) and standard dipole (bottom panel) at $z=0.25$ and $z=1$. The solid curves are our approximate expressions eqs.~\eqref{xirelcos} and~\eqref{stdip} and the dotted curves are the full-sky expressions. We see that the approximations recover well the full-sky expressions at sufficiently small separations, but are not accurate at very large separations. Up to~200$\,  h^{-1}$Mpc, our approximations are however sufficient: the fractional difference for both the relativistic and standard terms is smaller than a percent at those scale at $z=1$ and a few percent at $z=0.25$. The approximations capture therefore very well the dipole behavior in the regime we are interested in.   

\section{Conclusion}

\label{sec:conclusion}

A lot of interest has been devoted recently to the study of
relativistic effects in large-scale structure~\cite{galaxies,
challinor, Yoo, roy, class, class2, lombriser, bruni, roy21cm, 21cm}. 
In~\cite{galaxies, challinor, Yoo} a general relativistic derivation of the clustering of galaxies has been presented,
showing that besides redshift-space distortions, two types of relativistic effects appear: effects that 
are suppressed by power of ${\cal H}/k$ with respect to the standard Newtonian terms, namely gravitational redshift and velocity terms; and effects that are suppressed by a double power $\left({\cal H}/k\right)^2$, namely gravitational potential terms, integrated Sachs-Wolfe and Shapiro time delay. These effects are expected to carry complementary information to the
standard density and redshift-space distortions and may be useful to
test dark energy and modified gravity theories. 

In this paper, we concentrate on the first kind of effect, that are more likely to be detected first, and we investigate how they can be isolated.
The key point is to focus on the cross-correlation between two
different populations of galaxies, and 
make use of the different symmetries between the standard Newtonian terms and the relativistic terms to disentangle them. If we ignore the time evolution of the density and velocity perturbations, then in the flat-sky approximation the standard terms are purely symmetric whereas the relativistic terms are purely anti-symmetric (up to contributions of the order $\HH^2/k^2$, that are subdominant). By performing a multipole expansion of the two-points correlation function, we can therefore unambiguously separate the relativistic terms from the standard terms. 
Note that a similar multipole expansion of the relativistic terms has been investigated in Fourier space by McDonald~\cite{mcdonald} and Yoo et. al~\cite{uros}, and simulated by Croft~\cite{croft} (see also~\cite{roy} for the case of one population of galaxies).

In this paper, we show furthermore that the time evolution of the density and velocity perturbations and the wide-angle effects complicate the splitting: taking those into account we have found that the standard Newtonian terms also contribute to the anti-symmetric correlation function. However this contribution is smaller than the relativistic contribution and we have proposed a method to measure and remove the majority of this contamination. The anti-symmetric correlation function, and more precisely its dipolar modulation, seems therefore very promising to measure relativistic effects in large-scale structure. In a forthcoming paper we will apply our method to SDSS data.   

One can then wonder how these relativistic effects can be useful in
cosmology and what can be learned from them. One potential application
would be to use them to test dark energy and modified theories of
gravity. For example, we have seen that in theories of gravity where
the Euler equation holds (i.e. galaxies move on geodesics: they
respect the equivalence principle), the contribution from
gravitational redshift cancels out with the light-cone effect and part
of the Doppler effect. This needs not be the case in some classes of
modified gravity models \cite{HNS2009}. 
A measurement of the relativistic effects
can therefore be used to perform consistency
checks and to test for deviations from the standard paradigm of
general relativity $+$ $\Lambda$CDM.

\acknowledgments{ 
CB thanks Alex Hall for interesting discussions, Francesco Montanari for useful comparisons of the codes and Marc-Olivier Bettler for his help with the figures. 
CB is supported by the Herchel Smith Postdoctoral Fund and by King's College Cambridge. LH is supported in part by
the DOE under contract DE-FG02-92-ER40699 and NASA under contract
NNX10AH14G. 
LH also thanks Gary Shiu and Henry Tye at the IAS at the Hong Kong
University of Science and Technology for hospitality.
EG
acknowledges  support  by project AYA2009-13936, Consolider-Ingenio
CSD2007- 00060,   European Commission ITN CosmoComp
(PITN-GA-2009-238356) and  research project 2009- SGR-1398 
from Generalitat de Catalunya}

\appendix

\onecolumngrid

\section{Summary of the fully general relativistic calculation of $\Delta_{\rm obs}$}

\label{app:summary}

We present here a summary of the calculation of $\Delta_{\rm obs}$ based on ref.~\cite{galaxies} (note that here $\bn$ is the direction of observation, which points in the opposite direction as the $\bn$ used in ref.~\cite{galaxies} which denotes the photon direction). 
We start by calculating the redshift density perturbation $\delta_z$, defined as
\bea
\delta_z(z,\bn) &=& \frac{\rho(z,\bn)-\langle\rho\rangle(z)}{\langle\rho\rangle(z)}
 =\frac{\frac{N(z,\bn)}{V(z,\bn)}-\frac{\langle N\rangle(z)}{V(z)}}
{\frac{\langle N\rangle(z)}{V(z)}}
 = \frac{N(z,\bn)-\langle N\rangle(z)}{\langle N\rangle(z)}-\frac{\de
 V(z,\bn)}{V(z)}\, ,
\eea
where $\langle \rangle$ denotes the average over the direction $\bn$. 
The observed galaxy overdensity can therefore be expressed in terms of the redshift density perturbation $\delta_z$ and the volume perturbation
\be
\label{Npert}
\Delta_{\rm obs}(z, \bn)=\frac{N(z,\bn)-\langle N\rangle(z)}{\langle N\rangle(z)}=\de_z(z,\bn)+\frac{\delta
 V(z,\bn)}{V(z)}\, .
 \ee
 
The redshift density perturbation can be related to the density contrast by 
\bea
\de_z(z, \bn)=\frac{\rho(z, \bn)-\bar \rho(z)}{\bar \rho(z)}=
\frac{\bar{\rho}(\bar{z})+\delta\rho(z, \bn)-\bar\rho(z)}{\bar\rho(z)}
=\frac{\bar{\rho}(z-\de z)+\delta\rho(z,\bn)-\bar\rho(z)}{\bar\rho(z)}
=\frac{\de\rho(z, \bn)}{\bar\rho(\bar z)}-\frac{d\bar \rho}{d \bar z}
\frac{\delta z(z, \bn)}{\bar\rho(\bar z)}\, , 
\eea
where $\bar z =1/a(\eta)-1$ is the background redshift of a Friedmann universe and $\de z$ is the redshift perturbation.
The perturbation $\delta z$ can be calculated at linear order in the metric potentials and $\de_z(z, \bn)$ becomes
\begin{align}
&\de_z(z, \bn)=b\cdot\delta -3\bV\cdot\bn +3\Psi -3\frac{\HH}{k}V+3\int_{0}^{r} dr' (\dot{\Phi}+\dot{\Psi})\, .\label{dez}
\end{align}
Here $\delta$ is the density contrast in the comoving gauge and $b$ is the bias.

The volume element can be expressed as (for more details see~\cite{galaxies})
\begin{align}
&\frac{\de V}{V}(z, \bn) =    
 -3\Phi+\left( \cot\theta_O+\frac{\dd}{\dd \theta}\right)\de \theta +
\frac{\dd \de \varphi}{\dd \varphi} + \bV\cdot \bn+\frac{2\de r}{r}
-\frac{d\de r}{d\lambda}+\frac{1}{\HH(1+ z)}\frac{d \de z}{d\lambda}
-\left(\frac{2}{ r\HH}+\frac{\dot{\HH}}{\HH^2}-4 \right)\frac{\de z}{1+ z}\, ,
\end{align}
where $\lambda$ is the affine parameter of the geodesic, $\delta r$ denotes the radial perturbation along the geodesic and $\delta\theta$, $\delta\varphi$ are the transverse geodesic perturbations. These perturbations can be calculated by solving the null geodesic equation for the perturbed metric eq.~\eqref{metric}. With this we find for the volume perturbation
\begin{align}
&\frac{\de V}{V}=-2(\Psi+\Phi) + 4\bV\cdot\bn +\frac{1}{\HH}
\left[\dot\Phi+\dd_r\Psi-\frac{d(\bV\cdot\bn)}{dr}\right] 
+\left(\frac{\dot{\HH}}{\HH^2}+\frac{2}{r\HH}\right)
\left(\Psi-\bV\cdot\bn+ \int_{0}^{r} dr'(\dot{\Phi}+\dot{\Psi})\right)\nonumber\\
&-3\int_{0}^{r} dr'(\dot{\Phi}+\dot{\Psi})+ \frac{2}{r}\int_{0}^{r} dr' (\Phi+\Psi)
- \frac{1}{r}\int_{0}^{r} dr'\frac{r-r'}{r'}
\Delta_\Omega(\Phi+\Psi)~, \label{volume}
\end{align}
where $\Delta_\Omega$ denotes the Laplacian transverse to the line-of-sight.

Combining Eq.~\eqref{dez} with ~\eqref{volume} the observed overdensity of galaxies reads
\begin{align}
\Delta_{\rm obs}(z,\bn) &= b\cdot\delta -\frac{1}{\HH} \dd_r(\bV\cdot\bn)+\frac{1}{\HH}\dot{\bV}\cdot\bn
+\left(1-\frac{\dot{\HH}}{\HH^2}-\frac{2}{r\HH} \right)\bV\cdot\bn +\frac{1}{\HH}\dd_r\Psi - \frac{1}{r}\int_{0}^{r} dr'\frac{r-r'}{r'}
\Delta_\Omega(\Phi+\Psi)\nonumber\\
&+\left(\frac{\dot{\HH}}{\HH^2}+\frac{2}{r\HH} \right)\left[\Psi+\int_0^{r}dr'(\dot{\Phi}+\dot{\Psi})\right]
+\frac{2}{r}\int_{0}^{r} dr' (\Phi+\Psi) +\Psi-2\Phi+\frac{1}{\HH}\dot{\Phi}-3\frac{\HH}{k}V\, . \label{deltaobsapp}
\end{align}

This is consistent with  Eq. (\ref{Deltafin}) to $O({\cal H}/k)$,
setting $s = 0$.

\section{Details of the derivation of the relativistic two-point correlation function}
\label{app:derivationrel}

We want to calculate the relativistic correlation function
\be
\xirel(z, z', \theta)=\langle \Dst_\B(z,\bn)\Drel_\F(z',\bn') \rangle\, .
\ee

We use the Fourier convention
\be
f(\bx, \eta)=\frac{1}{(2\pi)^3}\int d^3k e^{-i\bk\bx}f(\bk,\eta)\, ,
\ee
and we relate the density, velocity and Bardeen potentials to the initial metric perturbation $\Psi_{\rm in}(\bk)$ via the transfer functions
\bea
D(\bk,\eta)&=&T_D(k,\eta)\Psi_{\rm in}(\bk)\, ,\\
V(\bk,\eta)&=&T_V(k,\eta)\Psi_{\rm in}(\bk)\, ,\\
\Psi(\bk,\eta)&=&T_\Psi(k,\eta)\Psi_{\rm in}(\bk)\, ,\\
\Phi(\bk,\eta)&=&T_\Phi(k,\eta)\Psi_{\rm in}(\bk)\, .
\eea
In General Relativity, if we neglect the anisotropic stress of the neutrinos, these functions become
\bea
T_\Phi&=&T_\Psi\, ,\\
T_D&=&-\frac{2a}{3\Omega_m}\left(\frac{k}{\HH_0}\right)^2 T_\Psi\, ,\\
T_V&=&\frac{2a\HH}{3\Omega_m\HH_0}\frac{k}{\HH_0}\left(T_\Psi+\frac{1}{\HH}\dot{T}_\Psi \right)\, .
\eea
Following~\cite{dod}, we furthermore decompose $T_\Psi(k,\eta)$ into a growth rate $D_1(a)$ and a time-independent transfer function\footnote{Note that we include here the pre-factor $9/10$ of~\cite{dod} in the transfer function $T(k)$.}~$T(k)$
\be
T_\Psi(k,\eta)=\frac{D_1(a)}{a}T(k)\, .
\ee 
The initial power spectrum is characterized by an amplitude $A$ and a scalar spectral index $n_s$
\be
k^3\langle\Psi_{\rm in}(\bk)\Psi_{\rm in}(\bk') \rangle=(2\pi)^3 A (k\eta_0)^{n_s-1} \delta(\bk+\bk')\, .
\ee 
With these definitions, the relativistic correlation functions becomes
\begin{align}
\xirel&(r,r',\theta)=A\int \frac{d^3k}{(2\pi)^3}e^{i\bk(\bx'-\bx)}\frac{(k\eta_0)^{n_s-1}}{k^3}\\
&\Bigg\{\left[\frac{\dot{\HH}(r')}{\HH^2(r')}+\frac{2}{r'\HH(r')} +5s_\B(r')\left(1-\frac{1}{r'\HH(r')} \right)\right]\cdot i(\hk\cdot\bn') T_V(k,r') \left(b_\B T_D(k,r)-\frac{k}{\HH(r)}(\hk\cdot\bn)^2T_V(k,r) \right) \nonumber\\
&-\left[\frac{\dot{\HH}(r)}{\HH^2(r)}+\frac{2}{r\HH(r)}+5s_\B(r)\left(1-\frac{1}{r\HH(r)} \right)\right]\cdot i(\hk\cdot\bn) T_V(k,r) \left(\!b_\F T_D(k,r')-\frac{k}{\HH(r')}(\hk\cdot\bn')^2T_V(k,r') \right) \Bigg\}\, .\nonumber
\end{align}
Following~\cite{szalay, szapudi, szapudi2, francesco}, we expand the exponential and the powers of $\hat{\bk}\cdot\bn$ (and similarly $\hat{\bk}\cdot\bn'$) in terms of spherical harmonics
\begin{align}
&e^{i\bk(\bx'-\bx)}=e^{id\bk\cdot\bN}=4\pi\sum_{LM}i^L j_L(k d)Y^*_{LM}(\hk)Y_{LM}(\bN)\, ,\nonumber\\
&\hk\cdot\bn=P_1(\hk \cdot\bn)= \frac{4\pi}{3}\sum_{m=-1}^1Y^*_{1m}(\hk)Y_{1m}(\bn)\, ,\nonumber\\
&(\hk\cdot\bn)^2=\frac{2}{3}P_2(\hk\cdot\bn)+\frac{1}{3}= \frac{8\pi}{15}\sum_{m=-2}^{2}Y^*_{2m}(\bn)Y_{2m}(\hk)+\frac{1}{3}\, .\nonumber
\end{align}
We can then perform the integrals over the direction of $\bk$. Terms with only one spherical harmonic enforce $L$ and $M$ to be zero:
\be
\int d\Omega_\bk  Y^*_{LM}(\hk)=\sqrt{4\pi}\delta_{L0}\delta_{M0}~.
\ee 
Terms with 2 spherical harmonics give rise to delta functions, e.g.
\be
\int d\Omega_\bk  Y^*_{LM}(\hk) Y_{2m}(\hk)=\delta_{L2}\delta_{Mm}\, ,
\ee
and terms with three spherical harmonics give rise to 3J Wigner symbols
\begin{align}
&\int d\Omega_\bk  Y^*_{LM}(\hk) Y^*_{2m}(\hk) Y^*_{1m'}(\hk)=\sqrt{\frac{(2L+1)5\cdot3}{4\pi}}
\begin{pmatrix}L & 2 & 1\\ -M & -m & -m'\end{pmatrix}\begin{pmatrix}L & 2 & 1\\ 0 & 0 & 0\end{pmatrix}\, .
\end{align}
Due to the properties of the Wigner symbols, only a finite number of terms contribute to the correlation function. The remaining spherical harmonics that depend on $\bn, \bn'$ and $\bN$ take a simple form in the coordinate system of Fig.~\ref{fig:coordinate}. With this the correlation function becomes
\begin{align}
\xirel=\frac{2A}{9\Omega_m^2\pi^2}\Bigg\{&R_1 \cos(\alpha)+R_2\cos(\beta)
+R_3\cos(\alpha)\cos(2\beta)+R_4\cos(\beta)\cos(2\alpha)\label{xirelfull}\\
&+R_5\sin(\alpha)\sin(2\beta)+R_6\sin(\beta)\sin(2\alpha)\Bigg\}\, .\nonumber
\end{align}
The coefficients $R_i$ depend on $r$, $r'$ and $d$. They read
\begin{align}
R_1(r,r',d)&=G_\F(r')D_1(r)D_1(r')f(r')
\left[\left(\bB(r)+\frac{2}{5}f(r) \right)\nu_1(d)-\frac{1}{10}f(r)\nu_3(d) \right]\\
R_2(r,r',d)&=-G_\B(r)D_1(r)D_1(r')f(r)
\left[\left(\bF(r')+\frac{2}{5}f(r') \right)\nu_1(d)-\frac{1}{10}f(r')\nu_3(d) \right]\\
R_3(r,r',d)&=G_\F(r')D_1(r)D_1(r')f(r)f(r')\frac{1}{5}\left[\nu_1(d)-\frac{3}{2}\nu_3(d) \right]\\
R_4(r,r',d)&=-G_\B(r)D_1(r)D_1(r')f(r)f(r')\frac{1}{5}\left[\nu_1(d)-\frac{3}{2}\nu_3(d)\right]\\
R_5(r,r',d)&=G_\F(r')D_1(r)D_1(r')f(r)f(r')\frac{1}{5}\Big[\nu_1(d)+\nu_3(d) \Big]\\
R_6(r,r',d)&=-G_\B(r)D_1(r)D_1(r')f(r)f(r')\frac{1}{5}\Big[\nu_1(d)+\nu_3(d)\Big]\, ,
\end{align}
where $D_1$ is the linear growth factor, and $f=\frac{d\ln D_1}{d \ln a}$. The functions $G_\F$ and $G_\B$ are defined as
\begin{align}
G_\F(r') &= \left[\frac{\dot{\HH}(r')}{\HH^2(r')}+\frac{2}{r'\HH(r')}+5s_\F(r')\left(1-\frac{1}{r'\HH(r')}\right)\right]\frac{\HH(r')}{\HH_0}\\
G_\B(r) &= \left[\frac{\dot{\HH}(r)}{\HH^2(r)}+\frac{2}{r\HH(r)}+5s_\B(r)\left(1-\frac{1}{r\HH(r)}\right)\right]\frac{\HH(r)}{\HH_0}
\end{align}
and
\begin{align}
\nu_\ell(d)&=\int\frac{dk}{k}\left(\frac{k}{H_0}\right)^3(k\eta_0)^{n_s-1}\,T^2(k)j_\ell(kd)\,, \hspace{0.1cm} \ell=1,3\, .
\end{align}

\begin{figure}[t]
\epsfig{figure=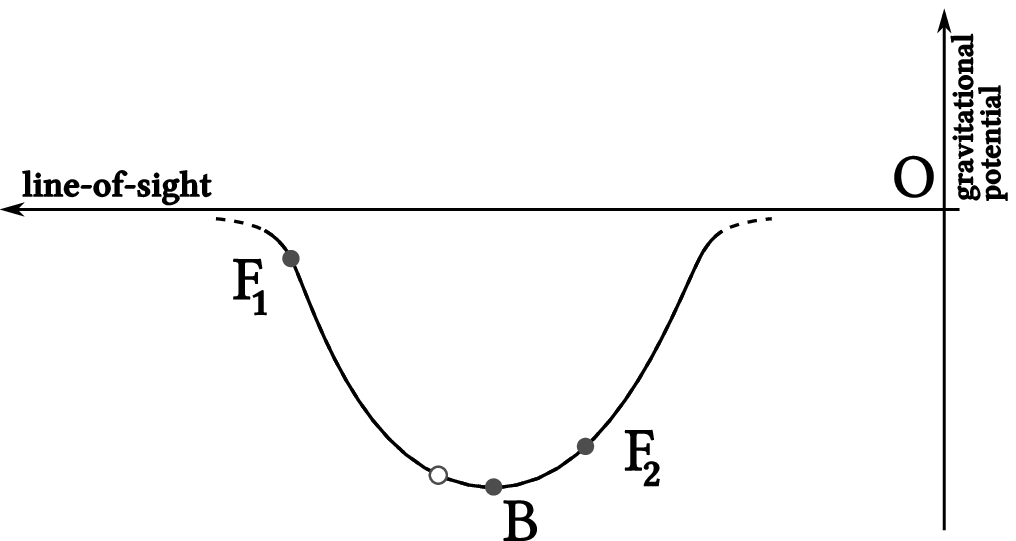,height=4cm}\hspace{2cm}\epsfig{figure=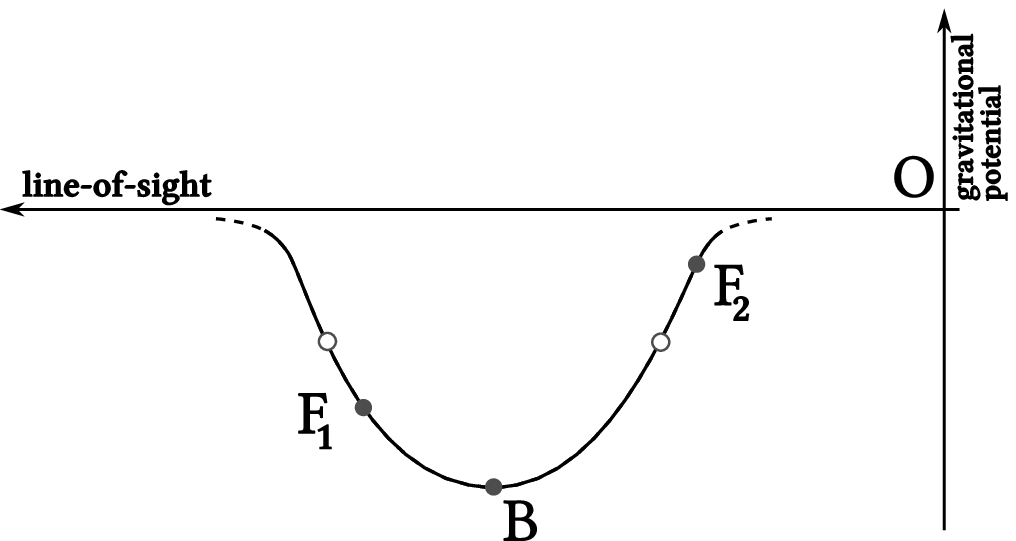,height=4cm}
\caption{ \label{fig:shiftBF} Sketch of the gravitational redshift effect. {\it Left panel}: we assume that only the bright galaxy is affected by gravitational redshift. The true position of B is denoted by a filled circle. Its observed position (open circle) is redshifted. The faint galaxies $\F_1$ and $\F_2$ are (by construction) symmetrically distributed around the observed position of the bright. This is because we are interested in comparing the observed correlation behind and in front of B, separated by the same redshift
separation. In real space, the faint galaxy in front, $\F_2$, is therefore closer to B than the faint galaxy behind, $\F_1$. The cross-correlation is consequently stronger for galaxies in front than behind the bright one. {\it Right panel}: we assume that only the faint galaxies are affected by gravitational redshift. The observed position of the faint galaxies (open circles) are by construction symmetrically distributed around the bright galaxy. Since $\F_1$ and $\F_2$ are both redshifted, in real space (filled circles) the faint galaxy behind, $\F_1$ is closer to B than the faint galaxy in front, $\F_2$. The cross-correlation is therefore stronger for galaxies behind than in front of the bright one. In reality, we measure a combination of the right and left panel. Since the right panel is weighted by the bias of the bright, $b_\B$, its contribution dominates over the left panel which is weighted by the bias of the faint, $b_\F<b_\B$. The total cross-correlation is therefore stronger for faint galaxies behind that in front of the bright.}
\end{figure}

In the flat sky approximation, eq.~\eqref{xirelfull} reduces to eq.~\eqref{xirelcos}.
In the absence of magnification bias, $s_\B=s_\F=0$, the sign of the correlation function~\eqref{xirelcos} is governed by the bias difference $b_\B-b_\F>0$. Two competing effects contribute to the cross-correlation: first the gravitational redshift of the bright galaxies modulated by the density of the faint galaxies, and second the gravitational redshift of the faint galaxies modulated by the density of the bright galaxies. As shown in Fig.~\ref{fig:shiftBF}, the first effect (left panel) generates a correlation which is stronger for faint galaxies in front of the bright galaxy (negative dipole), whereas the second effect (right panel) generates a correlation which is stronger for faint galaxies behind the bright galaxy (positive dipole). Since the bias of bright galaxies is larger than the bias of faint galaxies ($b_\B>b_\F$), the second effect dominates over the first one and the relativistic dipole is positive.

\section{Explicit expression for the multipolar coefficients of the standard contribution $S_i$}
\label{app:coeffst}

The standard coefficients of eq.~\eqref{xist} are
\begin{align}
S_1(r,r',d)=&D_1(r)D_1(r')\Bigg[\left(\bB(r)\bF(r')+\frac{\bB(r)}{3}f(r')+\frac{\bF(r')}{3}f(r)+\frac{2}{15}f(r)f(r')\right)\mu_0(d)\\
&-\frac{1}{3}\left(\frac{\bB(r)}{2}f(r')+\frac{\bF(r')}{2}f(r)+\frac{2}{7}f(r)f(r') \right)\mu_2(d) +\frac{3}{140}f(r)f(r')\mu_4(d)\Bigg]\nonumber\\
S_2(r,r',d)=&-D_1(r)D_1(r')\Bigg[\left(\frac{\bF(r')}{2}f(r)+\frac{3}{14}f(r)f(r') \right)\mu_2(d) +\frac{1}{28}f(r)f(r')\mu_4(d)\Bigg]\\
S_3(r,r',d)=&-D_1(r)D_1(r')\Bigg[\left(\frac{\bB(r)}{2}f(r')+\frac{3}{14}f(r)f(r') \right)\mu_2(d) +\frac{1}{28}f(r)f(r')\mu_4(d)\Bigg]\\
S_4(r,r',d)=&D_1(r)D_1(r')f(r)f(r')\left[\frac{1}{15}\mu_0(d)-\frac{1}{21}\mu_2(d)+\frac{19}{140}\mu_4(d) \right]\\
S_5(r,r',d)=&D_1(r)D_1(r')f(r)f(r')\left[\frac{1}{15}\mu_0(d)-\frac{1}{21}\mu_2(d)-\frac{4}{35}\mu_4(d) \right]
\end{align}
where
\begin{align}
\mu_\ell(d)&=\int\frac{dk}{k}\left(\frac{k}{H_0}\right)^4(k\eta_0)^{n_s-1}\,T^2(k)j_\ell(kd)\,, \hspace{0.1cm} \ell=0,2,4\,.\nonumber
\end{align}

\section{Explicit calculation of some of the Alcock-Paczynski contributions}

\label{app:AP}

Let us look at the Alcock-Paczynski contribution generated by the density term $\Delta(r,\bn)=b(\eta)\delta(r,\bn)$.

\subsection{Radial derivatives}

The density contribution in eq.~\eqref{xiAPradial} is
\begin{align}
\label{densAPradial}
\xiAP_\A=&\frac{r}{2}\Big[b_\B(\eta)b_\F(\eta')-b_\B(\eta')b_\F(\eta)\Big]\big(\dd_r+\dd_{r'} \big)\langle \delta(r,\bn)\delta(r',\bn')\rangle
\,\frac{\dd\ln r}{\dd\ln\Omega}\cdot \frac{\delta\Omega}{\Omega}\\
&- \frac{r}{2}\Big[b_\B(\eta)b_\F(\eta')\rangle
-b_\B(\eta')b_\F(\eta)\Big]\dd_{r'}\langle \delta(r,\bn)\delta(r',\bn')\rangle\frac{d}{r}\,\cos\beta\,\frac{\dd\ln\HH}{\dd\ln\Omega}\cdot\frac{\delta\Omega}{\Omega}\,  ,\nonumber
\end{align}
where the partial radial derivatives do not act on the bias $b_\B$ and $b_\F$ that are assumed to be scale independent.  
Expanding $b_\B(\eta')$ and $b_\F(\eta')$ around $\eta$, and using eq.~\eqref{cos} to relate $\cos\alpha$ to $\cos\beta$, eq.~\eqref{densAPradial} becomes
\begin{align}
\label{APradialfin}
\xiAP_\A(r,d, \beta)=&\frac{A}{9\Omega_m^2\pi^2}D_1^2 \,r(b_\B b'_\F-b'_\B b_\F) \int\frac{dk}{k}\left(\frac{k}{H_0}\right)^5(k\eta_0)^{n_s-1}T^2(k)j_1(kd)\\
& \cdot rH_0\left(\frac{d}{r}\right)^2\left[\frac{2}{5}\big(P_3(\cos\beta)-P_1(\cos\beta) \big)\frac{\dd\ln r}{\dd\ln\Omega}
+\frac{1}{5}\big(2P_3(\cos\beta)+3P_1(\cos\beta) \big)\frac{\dd\ln\HH}{\dd\ln\Omega} \right]\frac{\delta\Omega}{\Omega}\nonumber
\end{align}
Comparing this with eq.~\eqref{stdip} and~\eqref{stoct}, we see that here there is an additional factor $k/H_0$ which is compensated by the extra factor $d/r$ (note that the factor $rH_0$ at the beginning of the second line is of order 1). Consequently, eq.~\eqref{APradialfin} is suppressed by an overall factor $\delta\Omega/\Omega$ with respect to  eq.~\eqref{stdip} and~\eqref{stoct}.

\subsection{Time derivatives}

The density contribution in eq.~\eqref{xiAPtime} is
\begin{align}
\xiAP_\A=&-\frac{1}{2}\big(\dd_\eta+\dd_{\eta'} \big)\Big[\Big(b_\B(\eta)b_\F(\eta')-b_\B(\eta')b_\F(\eta)\Big)\langle \delta(r,\bn)\delta(r',\bn')\rangle\Big]
\frac{\dd\ln r}{\dd\ln\Omega}\cdot \frac{\delta\Omega}{\Omega}\nonumber\\
&+ \frac{1}{2}\dd_{\eta'}\Big[\Big(b_\B(\eta)b_\F(\eta')-b_\B(\eta')b_\F(\eta)\Big)\langle \delta(r,\bn)\delta(r',\bn')\rangle\Big]
\frac{d}{r}\,\cos\beta\,\frac{\dd\ln\HH}{\dd\ln\Omega}\cdot\frac{\delta\Omega}{\Omega}\,  ,\nonumber
\end{align}
where the time derivatives act both on the bias and the density contrast $\delta(r,\bn)$. Expanding $b_\B(\eta')$ and $b_\F(\eta')$ and their time derivatives $\partial_{\eta'}b_\B(\eta')$ and $\partial_{\eta'}b_\F(\eta')$ around $\eta$, and using eq.~\eqref{cos} to relate $\cos\alpha$ to $\cos\beta$, we find
\begin{align}
\label{APtimefin}
\xiAP_\A(r,d, \beta)&=\frac{A}{9\Omega_m^2\pi^2}  \int\frac{dk}{k}\left(\frac{k}{H_0}\right)^4(k\eta_0)^{n_s-1}T^2(k)j_0(kd)\\
&\Bigg\{
r^2\Big[\big(b_\B b''_\F-b_\F b''_\B \big)D_1^2+2\big(b_\B b'_\F-b_\F b'_\B \big)D_1 D'_1 \Big]\frac{\dd\ln r}{\dd\ln\Omega}-r \big(b_\B b'_\F-b'_\B b_\F\big)D_1^2\, \frac{\dd\ln\HH}{\dd\ln\Omega}\Bigg\} \frac{d}{r} P_1(\cos\beta) \frac{\delta\Omega}{\Omega}\, .\nonumber
\end{align}
Here $rD'_1=-r\HH f D_1$ is of the order of $f D_1$ and $r^2b''$ is roughly of the order of $r^2\HH^2 d^2b/dz^2\sim d^2b/dz^2$.
Consequently eq.~\eqref{APtimefin} is similar to eq.~\eqref{stdip} and~\eqref{stoct}, apart from an overall suppression from the factor $\delta\Omega/\Omega$.

\section{Comparison with cluster measurements}

\label{app:cluster}

Let us compare our approach with cluster
measurements, as in the case of WHH \cite{nature}.
WHH measures a mean redshift difference between the
brightest central galaxy and the rest of the cluster galaxies.
The estimator can be thought of as:
\be
\langle z_\F-z_\B\rangle=\sum_{i,j} \frac{1}{\bar{n}^\F \bar{n}^\B} (z_i-z_j)n_i^\F n_j^\B=\sum_{i,j}  (z_i-z_j)\left(1+\delta_i^\F\right)\left(1+\delta_j^\B\right)
\, ,
\label{redshift}
\ee
where conceptually, we can think of the survey as being pixelized 
such that each pixel contains either one or zero galaxy,
with $n^\B_j$ denoting that central (brightest) galaxy number ($1$ or $0$)
in pixel $j$,
and $n_i^\F$ denoting the corresponding number in
pixel $i$ for the rest of the cluster members,
and $\bar n^\B$ and $\bar n^\F$ being their respective mean.
The summation over $i$ and $j$ over the whole survey is equivalent to
stacking many clusters to pull out a signal.

There are two, subtly different, ways to think about this estimator.
From one point of view (ours), redshifts are converted into distances
using some background cosmology.
Once that is done, $z_j$ and $z_i$ are merely coordinate labels,
and the factor of $(z_i - z_j)$ can
be pulled out of an ensemble average. 
Thus, the ensemble average of Eq. (\ref{redshift}) really
has to do with $\langle n_i^\F n^\B_j \rangle$; the sum over
$i$ and $j$ with the $(z_i - z_j)$ weighting serves to isolate
the anti-symmetric part of the cross-correlation function
between the central brightest galaxy and the rest of the cluster.
Hence, from our point of view, Eq. (\ref{redshift})
essentially measures the first moment of the cross-correlation
function.

There is another way to think about this estimator, however.
One can think of the redshift ($z_i$ and $z_j$) as a stochastic
quantity i.e. after all, it contains fluctuations due to Doppler
and gravitational redshifts.
From that point of view, an ensemble average of 
Eq. (\ref{redshift}), to the lowest order in fluctuations,
gives cross-correlations between fluctuations in redshift
to fluctuations in the galaxy density.
In this second point of view, one should think of the pixelization
as being done in the {\it true} ${\bf  x}$ space, as opposed to the
{\it apparent} ${\bf x}$ space (inferred using some background
cosmology). It is only in the former where one can meaningfully
talk about the redshift as a fluctuation variable (i.e. dependent on
the true ${\bf x}$).

The two approaches should be equivalent. However, translating our result in terms of redshift differences, we see that the sign of our average redshift difference at first sight seems inconsistent with the cluster result. We predict indeed a positive dipole: the correlation function is therefore stronger for faint galaxies behind the bright galaxy, which in turn results in a net {\it redshift} of the faint galaxies with respect to the bright one
\be
\label{xidifflss}
\langle z_\F-z_\B\rangle=\int_{z_\B-\Delta z}^{z_\B+\Delta z}\!dz'_\F\,(z'_\F-z_\B)\xi(z'_\F,z_\B)
=2 \int_{z_\B}^{z_\B+\Delta z}\!dz'_\F\,(z'_\F-z_\B)\xi_\A(z'_\F,z_\B)\, ,
\ee
for a chosen redshift range $\Delta z$ and a fixed value of $z_\B$~\footnote{Note that stacking different clusters simply adds an additional average over the redshift of the bright galaxy, $z_\B$, in eq.~\eqref{xidifflss}.}. In the second equality we have used that the symmetric part of the correlation function vanishes due to the anti-symmetry of $z'_\F-z_\B$ and to the symmetric boundaries  of integration. Since $\xi_\A>0$ for $z'_\F>z_\B$, the right-hand side of eq.~\eqref{xidifflss} is manifestly positive.

Cluster measurements on the other hand find a net {\it blueshift} of the faint galaxies with respect to the central brightest galaxy. This apparent inconsistency comes from the fact that, in the cluster measurements, the redshift difference is averaged over {\it all} galaxies that belong to the cluster
\be
\label{xidiffclust}
\langle z_\F-z_\B\rangle=\int_{z_\B-\Delta z_{\rm min}}^{z_\B+\Delta z_{\rm max}}\!dz'_\F\,(z'_\F-z_\B)\xi(z'_\F,z_\B)\, .
\ee
Here $\Delta z_{\rm min}$ and $\Delta z_{\rm max}$ denote the physical boundaries of the cluster. Since the brightest central galaxy is in average more redshifted than the other galaxies in the cluster, $\Delta z_{\rm min}>\Delta z_{\rm max}$. Eq.~\eqref{xidiffclust} becomes then
\be
\label{xiclusttot}
\langle z_\F-z_\B\rangle=2 \int_{z_\B}^{z_\B+\Delta z_{\rm max}}\!dz'_\F\,(z'_\F-z_\B)\xi_\A(z'_\F,z_\B)+
\int_{z_\B-\Delta z_{\rm min}}^{z_\B-\Delta z_{\rm max}}\!dz'_\F\,(z'_\F-z_\B)\xi(z'_\F,z_\B)\, .
\ee
The second term dominates over the first one since it depends on the total correlation function $\xi$ which is significantly larger than the anti-symmetric contribution $\xi_\A$. This second term is negative since within $z_\B-\Delta z_{\rm min}$ and $z_\B-\Delta z_{\rm max}$, $z'_\F-z_\B$ is negative and $\xi$ is positive. The net redshift difference is therefore negative in the case of the clusters, as found in~\cite{nature}. Eq.~\eqref{xiclusttot} clearly shows that what dominates in the cluster measurements is actually the gravitational redshift of the {\it boundaries} of the cluster, which in redshift space are asymmetrically distributed around the centre of the cluster. As such the cluster measurement is a measurement of the three-point correlation function between the density of the bright galaxies, the density of the faint galaxies and the gravitational redshift of the boundaries. On the other hand, in our case, from eq.~\eqref{xidifflss} we see that we really measure a two-point correlation function, because the range of integration is defined arbitrarily around the central galaxy and is by construction symmetric.


\end{document}